\documentclass[acmsmall]{acmart}

\setcopyright{cc}
\setcctype{by-nc}
\acmDOI{10.1145/3728972}
\acmYear{2025}
\acmJournal{PACMSE}
\acmVolume{2}
\acmNumber{ISSTA}
\acmArticle{ISSTA095}
\acmMonth{7}
\received{2025-02-27}
\received[accepted]{2025-03-31}

\AtBeginDocument{%
  }

\usepackage{amsmath}
\usepackage[ruled,vlined,linesnumbered,boxed,commentsnumbered]{algorithm2e}
\usepackage{multirow}
\usepackage{algorithmicx}
\usepackage{booktabs}
\usepackage[normalem]{ulem}
\usepackage{graphicx}
\usepackage{wrapfig}
\usepackage{makecell}
\usepackage{tablefootnote}
\usepackage{subcaption}
\usepackage{threeparttable}
\newcommand{\finding}[2]{
    \begin{center}
    \fcolorbox{black}{gray!10}{\parbox{.97\linewidth}{
    {#2}
    }}
    \end{center}
}

\newcommand{\myz}[1]{\textcolor{black}{#1}}
\newcommand{\tool}{ModelMeta }
\newcommand{\toolnospace}{ModelMeta}

\begin{document}


\title{Improving Deep Learning Framework Testing with Model-Level Metamorphic Testing}

\author{Yanzhou Mu}
\orcid{0000-0003-1816-2246}
\affiliation{%
  \institution{Nanjing University}
  \city{Nanjing}
  \country{China}
}
\affiliation{%
  \institution{Shenzhen Research Institute of Nanjing University}
  \city{Shenzhen}
  \country{China}
}
\email{602022320006@smail.nju.edu.cn}

\author{Juan Zhai}
\orcid{0000-0001-5017-8016}
\affiliation{%
  \institution{University of Massachusetts at Amherst}
  \city{Amherst}
  \country{USA}
}
\email{juanzhai@umass.edu}

\author{Chunrong Fang}
\authornote{Chunrong Fang is the corresponding author.}
\orcid{0000-0002-9930-7111}
\affiliation{%
  \institution{Nanjing University}
  \city{Nanjing}
  \country{China}
}
\affiliation{%
  \institution{Shenzhen Research Institute of Nanjing University}
  \city{Shenzhen}
  \country{China}
}
\email{fangchunrong@nju.edu.cn}

\author{Xiang Chen}
\orcid{0000-0002-1180-3891}
\affiliation{%
  \institution{Nantong University}
  \city{Nantong}
  \country{China}
}
\email{xchencs@ntu.edu.cn}

\author{Zhixiang Cao}
\orcid{0009-0002-6810-7608}
\affiliation{%
  \institution{Nantong University}
  \city{Nantong}
  \country{China}
}
\email{zxcao@stmail.ntu.edu.cn}

\author{Peiran Yang}
\orcid{0009-0008-8242-9543}
\affiliation{%
  \institution{Nanjing University}
  \city{Nanjing}
  \country{China}
}
\email{peiranyang@smail.nju.edu.cn}

\author{Kexin Zhao}
\orcid{0009-0008-7793-2507}
\affiliation{%
  \institution{Nanjing University}
  \city{Nanjing}
  \country{China}
}
\email{221250059@smail.nju.edu.cn}

\author{An Guo}
\orcid{0009-0005-8661-6133}
\affiliation{%
  \institution{Nanjing University}
  \city{Nanjing}
  \country{China}
}
\email{guoan218@smail.nju.edu.cn}

\author{Zhenyu Chen}
\orcid{0000-0002-9592-7022}
\affiliation{%
  \institution{Nanjing University}
  \city{Nanjing}
  \country{China}
}
\affiliation{%
  \institution{Shenzhen Research Institute of Nanjing University}
  \city{Shenzhen}
  \country{China}
}
\email{zychen@nju.edu.cn}

\renewcommand{\shortauthors}{Mu et al.}

\begin{abstract}
  Deep learning (DL) frameworks are essential to DL-based software systems, and framework bugs may lead to substantial disasters, thus requiring effective testing.
  Researchers adopt DL models or single interfaces as test inputs and analyze their execution results to detect bugs. 
  However, floating-point errors, inherent randomness, and the complexity of test inputs make it challenging to analyze execution results effectively, leading to existing methods suffering from a lack of suitable test oracles. 
  Some researchers utilize metamorphic testing to tackle this challenge.
  They design Metamorphic Relations (MRs) based on input data and parameter settings of a single framework interface to generate equivalent test inputs, ensuring consistent execution results between original and generated test inputs.
  Despite their promising effectiveness, they still face certain limitations. 
  (1) Existing MRs overlook structural complexity, limiting test input diversity. 
  (2) Existing MRs focus on limited interfaces, which limits generalization and necessitates additional adaptations.
  (3) Their detected bugs are related to the result consistency of single interfaces and far from those exposed in multi-interface \myz{combinations} and runtime metrics (\myz{e.g.,} resource usage).
  To address these limitations, we propose \toolnospace, a model-level metamorphic testing method for DL frameworks with four MRs focused on the structure characteristics of DL models. \tool augments seed models with diverse interface combinations to generate test inputs with consistent outputs, guided by the QR-DQN strategy. It then detects bugs through fine-grained analysis of training loss/gradients, memory/GPU usage, and execution time.
  We evaluate the effectiveness of \tool on three popular DL frameworks (i.e., MindSpore, PyTorch, and ONNX) with 17 DL models from ten real-world tasks ranging from image classification to object detection. 
  Results demonstrate that \tool outperforms state-of-the-art baselines from the perspective of test coverage and diversity of generated test inputs.
  Regarding bug detection, \tool has identified 31 new bugs, of which \myz{27} have been confirmed, and 11 have been fixed. Among them, seven bugs existing methods cannot detect, i.e., five \myz{wrong} resource usage bugs and two low-efficiency bugs. These results demonstrate the practicality of our method.\looseness=-1
\end{abstract}




\maketitle


\vspace{-2mm}
\section{Introduction}
\label{sec:intro}

Modern software systems powered by deep learning (DL) models have been widely applied in various industry domains, such as autonomous driving~\cite{ChenSKX15}, machine translation~\cite{gao2019soft}, medical diagnosis~\cite{obermeyer2016predicting}, and software engineering~\cite{
chen2021graph,chen2019continuous}. To enable efficient development, execution, and deployment of these systems, DL frameworks provide standard interfaces, optimize resource scheduling, and support compatibility migration for DL models. Therefore, the quality assurance of DL frameworks has attracted significant attention, as their bugs are often subtle but may cause harmful effects on DL models, even leading to serious real-world accidents~\cite{teslanews}. Until now, researchers have proposed different DL framework testing methods~\cite{pham2019cradle,2020audee,wang2020lemon,li2023comet,Deng2022FuzzingDL,2022metamorphic,chen2024miss,Wei2022FreeLF,zhang2021predoo,liu2023generation,shen2024tale}.
\looseness=-1

Researchers often use DL models~\cite{pham2019cradle,2020audee,wang2020lemon,li2023comet} or single framework interfaces~\cite{Deng2022FuzzingDL,xie2021leveraging,xie2024cedar,Wei2022FreeLF,zhang2021predoo} as inputs to test DL frameworks. 
However, floating-point errors~\cite{zhou2024polyjuice}, inherent randomness, and test input complexity (e.g., large-scale weights) complicate the analysis of execution behaviors for bug detection, resulting in a lack of reliable test oracles.
To alleviate this challenge, researchers~\cite{wang2022eagle,ding2017validating,chen2024miss} resort to metamorphic testing~\cite{chen1998metamorphic}. 
Specifically, these studies design Metamorphic Relations (MRs) based on the characteristics of single interfaces, including the input data or parameter settings, to generate new test inputs. 
The newly generated input should remain consistent with the original one, and any violation is deemed to expose bugs.
Based on MRs, these methods achieve test oracles and conduct certain bug detection.
However, they still have the following limitations. \looseness=-1

\textbf{Limitation 1: Shallow Metamorphic Relations.} 
They propose interface-level MRs based on input data (e.g., shape and dimension) and parameter settings (e.g., value type) but overlook the structure complexity of interface combinations, which often triggers bugs. For instance, a TensorFlow bug involving a model with ``Conv3D'' and ``SVD'' interfaces~\cite{bugcase} causes gradient calculations to return ``NaN'' in TensorFlow 2.5 and ``Segmentation Fault'' in TensorFlow 2.8. 
These errors stem from data interaction between multiple interfaces, though each interface can execute correctly and separately. \looseness=-1

\textbf{Limitation 2: Limited Generalization Ability.} 
Since existing MRs typically rely on parameters or input characteristics specific to certain interfaces, they only cover a limited subset of framework interfaces. Testing new interfaces often demands substantial manual effort to design new MRs, including analysis of parameter ranges and other properties. This constraint limits the generalization of MRs to new interfaces and reduces their practical applicability.


\looseness=-1

\textbf{Limitation 3: Restricted Detected Bug Types.} 
Existing methods have limited scope and can detect only a narrow range of bugs. They struggle to identify bugs that arise within groups of multiple interfaces since they primarily focus on detecting output inconsistencies and implementation error bugs at the single interface level, overlooking bugs related to abnormal runtime metrics (e.g., resource usage, and execution efficiency), which are also commonly encountered in real applications. 
\looseness=-1

To overcome the above limitations, we propose \toolnospace, a novel model-level metamorphic testing method for DL frameworks. It designs four MRs based on the structural characteristics of DL models to generate new equivalent DL models. These MRs are implemented as specific interface groups (e.g., complete square calculation as shown in the motivating example in Section~\ref{sec:mrs}) to ensure output consistency after inserting various external structures into the original model.
This makes the model structure more complex, while the new model should produce the same output as the original one. \tool can detect bugs by analyzing the generated models' outputs.
Specifically, our designed MRs have the following advantages.
First, these MRs are related to the structure characteristics, which can improve the diversity of various groups of different interfaces (addressing \textbf{Limitation 1}).
Second, common external structures of various interfaces or a single interface can be inserted to generate new models without designing new MRs or adaptations, as our MRs can ensure equivalence between the original and generated models regardless of the inserted structures (addressing \textbf{Limitation 2}).
Finally, by introducing diverse interface groups based on these MRs, \tool effectively detects bugs exposed in multiple interfaces. Moreover, it monitors and analyzes various runtime metrics, including resource usage, execution time, and training loss/gradients, enhancing its ability to detect new bugs (addressing \textbf{Limitation 3}). 
We further utilize guidance strategies to effectively generate more diverse and valuable test inputs (especially those that can trigger bugs), i.e.,  \tool adopts the QR-DQN strategy~\cite{dabney2018distributional}, combined with the $\epsilon$-greedy strategy~\cite{mnih2015human} and Thompson Sampling strategy~\cite{thompson1933likelihood} to select MRs and seed models, respectively. 
Besides, 
\tool also migrates interface MRs on the models after applying structure MRs to improve the diversity of generated test inputs. Finally, \tool analyzes the memory/GPU resource misuse, inconsistent output of loss/gradient, crashes, and execution time of the generated test inputs to detect more diverse bugs.

To evaluate \toolnospace, we conduct large-scale experiments on three popular DL frameworks: PyTorch~\cite{torch}, MindSpore~\cite{mindspore}, and ONNX~\cite{onnx}. We collect 17 popular DL models from ten industry tasks, ranging from image classification to semantic segmentation. These models include various interface combinations for different functionalities (e.g., residual structures~\cite{he2016deep} for image feature extraction). \tool detects 31 bugs; 27 have been confirmed, and 11 bugs have been fixed. Among the 31 detected bugs, \tool detects two new kinds of bugs that existing methods cannot detect, including five unreasonable resource usage bugs and two low-efficiency bugs. This demonstrates the effectiveness of \tool. \looseness=-1

In summary, the main contributions of our work are summarized as follows:

\vspace{-1mm}
\begin{itemize}
    
    \item \textbf{Innovative Approach.} We propose \toolnospace, the first model-level metamorphic testing method for DL frameworks. Based on our newly designed MRs, it can improve test input diversity by covering new interfaces and their groups without restrictions like interface type. Besides, it introduces the QR-DQN~\cite{dabney2018distributional} combined with the $\epsilon$-greedy~\cite{mnih2015human} and Thompson Sampling strategy~\cite{thompson1933likelihood} to guide the test input generation, which can effectively explore the test input space and trigger bugs.\looseness=-1

    \item \textbf{Comprehensive Evaluation.} Experimental results show promising performance of \tool in generating diverse test inputs and also excels in bug detection. It has detected 31 new bugs, with \myz{27} confirmed and 11 fixed. Besides, \tool can detect two new kinds of bugs, such as memory leaks and low-efficiency bugs.

    \item \textbf{Promising Practicality.} The executable version of \tool and all experimental results are available on our website~\cite{sharelink}. Researchers can run it on target DL frameworks to detect bugs, with easy modifications guided by our instructions.
    Furthermore, we apply \tool to the daily tasks of our industry partners and receive high recognition from them. \looseness=-1

\end{itemize}

\vspace{-2mm}
\section{Background}
\label{sec:background}

\subsection{Metamorphic Testing}
\label{sec:mrrelated}
Chen et al.~\cite{chen1998metamorphic} propose metamorphic testing (MT) to address the challenge of the lack of test oracles~\cite{barr2014oracle}, where constructing the expected output to verify the program's execution is complex. It generates new test inputs based on those passed test inputs that do not trigger bugs and further detects whether there are violations among the execution results between the new test inputs and the original ones based on the preset MRs, regardless of checking the correctness of each test input. MRs reflect the constraint relationships between the original and generated test inputs. 
If the results of the generated test inputs violate the MR, one bug is reported. For instance, the MR can be defined as $f(x) = -f(-x)$, where $x$ is the original test input, $-x$ is the generated new test input, and $f$ is the program to be tested. The constraint of the MR is the execution result of $x$, which must equal the opposite execution result of $-x$ in program $f$. \looseness=-1

\vspace{-3mm}
\subsection{Reinforcement Learning}
\label{sec:rllearning}
\myz{Reinforcement Learning (RL)~\cite{jay2019deep} is one type of machine learning algorithm where an agent learns to optimize its decision-making by interacting with the environment. A key feature of RL is its feedback loop, where the agent continuously updates its policy based on received rewards and observed state transitions, aiming to maximize cumulative future rewards. Such a process is formalized by the Markov Decision Process (MDP)~\cite{swiechowski2023monte}, which models the agent-environment interaction. MDP consists of four key components: \textbf{(1) State}, representing the environment's condition at a given time; \textbf{(2) Action,} defining the possible behaviors an agent can take in a given state to affect the environment and obtain rewards; \textbf{(3) Reward,} which provides feedback on the evaluation of an agent’s action; and \textbf{(4) Policy,} the strategy guiding the agent’s actions to maximize cumulative rewards over time. RL has proven effective in complex decision-making tasks and is widely applied in areas such as financial forecasting~\cite{moody2001learning}, game strategy optimization~\cite{silver2016mastering}, and network traffic management~\cite{mnih2016asynchronous}.}

\myz{As one popular RL method, the QR-DQN strategy integrates quantile regression~\cite{dabney2018distributional} with the DQN algorithm~\cite{mnih2013playing} to dynamically evaluate global rewards by learning value distributions. It remains effective with sparse or imperfect rewards, minimizing sensitivity to reward design. It accurately approximates values with limited early data, adapting and guiding action and state selections in complex, uncertain environments while dynamically adjusting the discount factor to address information scarcity during early exploration. }\looseness=-1

\vspace{-2mm}
\section{Approach}
\label{sec:approach}
\subsection{Overview}
\label{sec:overview}

\begin{figure}[]
     \centering
    \includegraphics[width=0.98\linewidth]{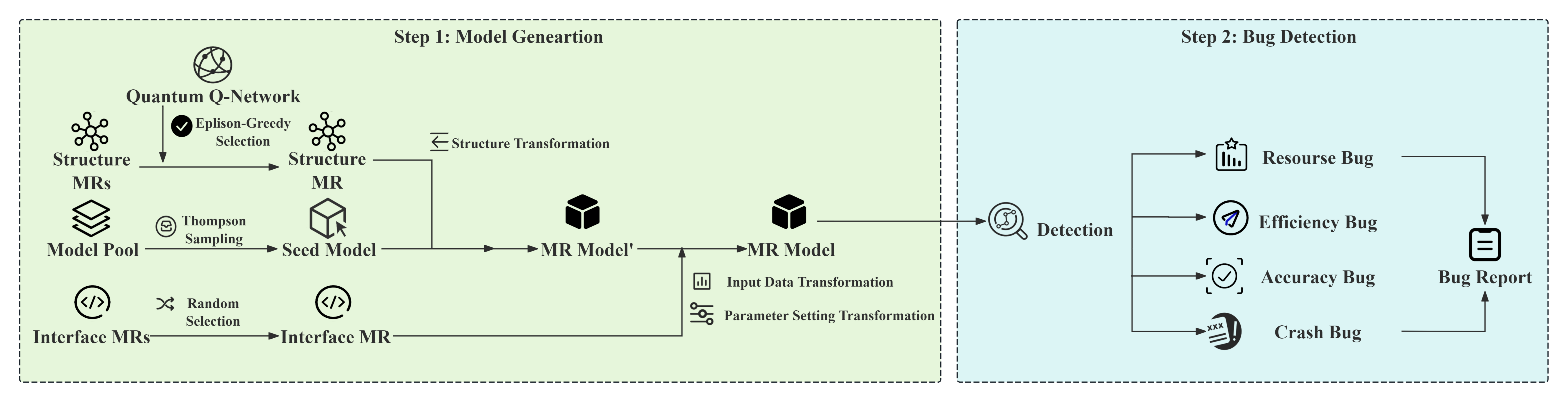}
    \vspace{-4mm}
     \caption{Workflow of \tool}   
     \label{fig:workflow}
\end{figure}

Our proposed method, \toolnospace, comprises two main components, as shown in Fig.~\ref{fig:workflow}: 
(1) \textit{Model Generation} and (2) \textit{Bug Detection}. 
During \textit{Model Generation}, \tool employs the QR-DQN algorithm~\cite{dabney2018distributional} combined with the $\epsilon$-greedy strategy~\cite{mnih2015human} to select structure MR (SMR) and generate a new model $MR Model'$. To further enhance the diversity of internal layers, \tool adopts random selection for the interface MR (IMR) and applies it on $MR Model'$ to generate $MR Model$. If it crashes, \tool uses the Thompson Sampling strategy~\cite{thompson1933likelihood} to select one new seed from the previously generated model pool for the next iteration.
The process finishes when the execution rounds reach the specified number and enter bug detection. 
During \textit{Bug Detection}, \tool analyzes all valid generated models to detect resource, efficiency, accuracy, and crash bugs by monitoring memory/GPU resource usage, execution time, model outputs, training loss, gradients, and execution logs.
In the rest of Section~\ref{sec:approach}, we first introduce the details of the adopted MRs (Section~\ref{sec:mrs}), then present the model generation process of \tool (Section~\ref{sec:seedselection}, Lines $1$-$22$ in Algorithm~\ref{alg:modelMeta}), and finally introduce the bug detection process, including the test oracles (Section~\ref{sec:bugdetection}, Lines $23$-$27$ in Algorithm~\ref{alg:modelMeta}). \looseness=-1

\begin{algorithm}[t]
	\scriptsize
	\SetAlgoVlined
	\SetKwInOut{Input}{\textbf{Input}}\SetKwInOut{Output}{\textbf{Output}}
	\Indm
	\Indp
	\Input{$M$: the original DL model; \
        $InsertPool$: the insert layer pool;\
        $\epsilon$, $\epsilon\_end$, $\epsilon\_decay$: the parameters that control the randomness degree during execution;\
        $N$: the number of execution rounds;\
        
		}
	\Output{$BugList$: List of all detected bugs;}
	\BlankLine
 
	$n \gets 0$, d $\gets M$, $BugList \gets \emptyset$  $ReplayPool$ $\gets \emptyset$, Init($QuantumQ$, $TargetQ$); //
  $QuantumQ$ is the quantum deep Q-network used for MRs selection, while $TargetQ$ is another quantum deep Q-network used for updating $QuantumQ$'s weights.
 \\
    \ForEach{n in $Range( 1, N)$}{
        $p$ $\gets$ $UniformSampling(0, 1)$;\\
        \eIf{$p$ $\leq \epsilon$}
        {
           $MR_{Structure}$ $\gets RandomSelect(MRPool_{structure})$;\\
           
        }
        {
            $MR_{Structure}$ $\gets MRPool_{Structure}[\arg \max (QuantumQ.calculate(d, *))]$;\\
        }
        $InsertLayer$ $\gets$ $RandomSelect(InsertPool)$; \\ $MR_{Interface}$ $\gets RandomSelect(MRPool_{Interface})$;
        \\
        $d'$ $\gets$ $MR_{Interface}.transform(MR_{Structure}.transform(d, InsertLayer))$;\\
        \eIf{Judge($d'$) == -1}{
		  	r $\gets$ -1, Done $\gets$ True; 
     \\
                $d'$ $\gets ThompsonSampling(ReplayPool)$;\\
		}
        {
            r $\gets$ $CalDiversity(d')$, Done $\gets$ False;\\
            $ReplayPool$.add(($d'$,  r));\\
        }
        
        $QuantumQvalue$ $\gets$ $QuantumQ.calculate(d, MR_{Structure})$;\\
        $TargetQvalue$ $\gets$ $r$+ $\gamma$ * $\max (TargetQ.calculate(d', *))$ * (1-Done);\\
        $Update(QuantumQ, TargetQvalue, QuantumQvalue)$;\\
		$TargetQ$ $\gets$ $QuantumQ.copy()$;\\  
        $\epsilon$ $\gets$ $\max(\epsilon_{end}, \epsilon_{decay} * \epsilon)$;\\  
        d $\gets$ d'.copy();\\
    }
    
    $LeftModels$ $\gets ReplayPool.getmodels()$;\\
       
       \ForEach{$d$ in $LeftModels$}{
           $NewBug \gets DetectBug$($d$);\\
           $BugList$.add($NewBug$); 
           \\
        }
    return $BugList$
	\caption{The \tool Algorithm}
 
	\label{alg:modelMeta}
\end{algorithm}

\subsection{Design of Metamorphic Relations}
\label{sec:mrs}
As shown in Algorithm~\ref{alg:modelMeta} (Line $10$), \tool first applies the selected SMR and then applies the selected IMR to generate new models. Next, we introduce the two kinds of MRs, respectively. \looseness=-1

\begin{figure}[htbp]
    \centering
     \begin{subfigure}[b]{0.49\textwidth}
        \includegraphics[width=\textwidth]{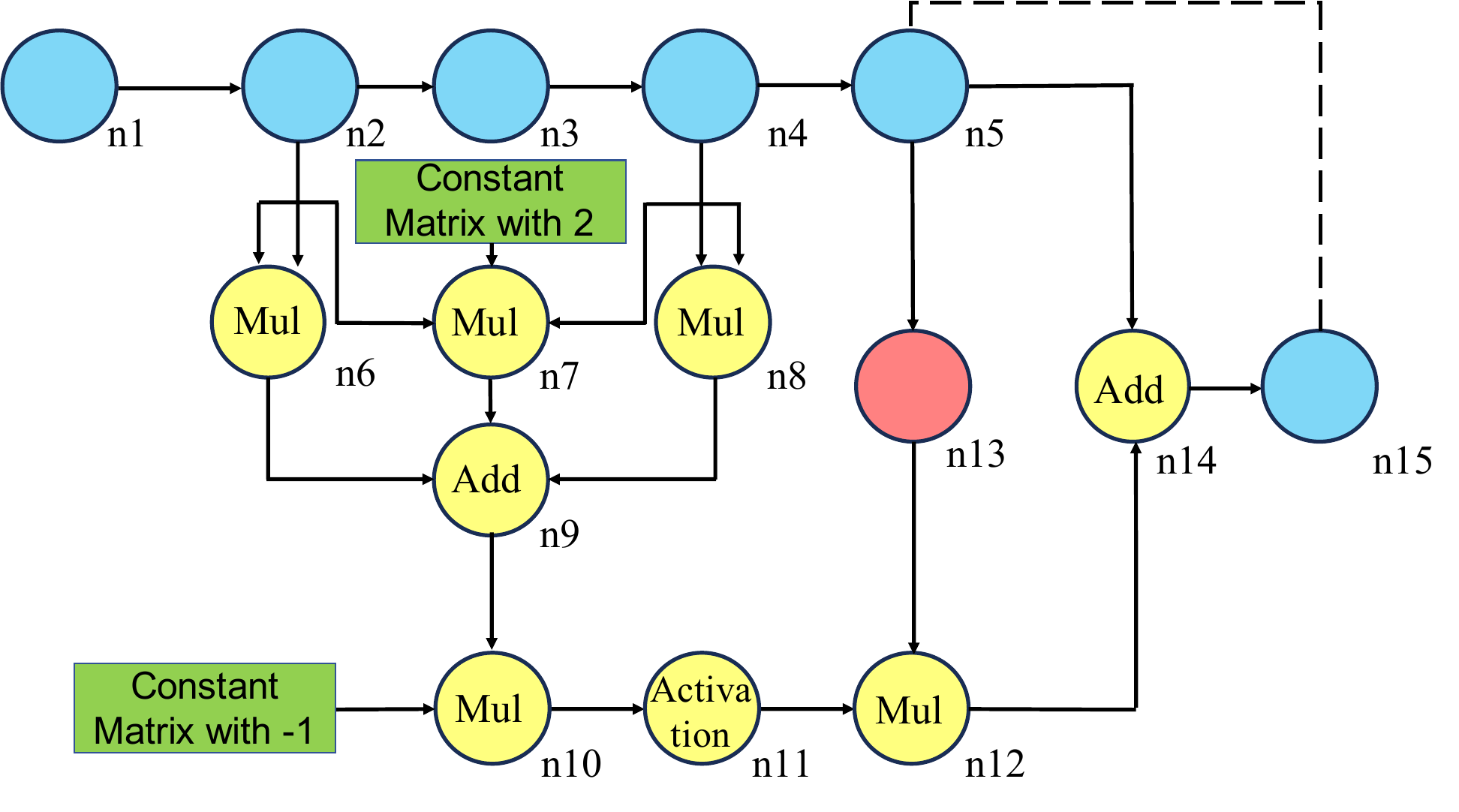}
        \caption{SMR1}
        \label{fig:smr1}
    \end{subfigure}
    \begin{subfigure}[b]{0.49\textwidth}
        \includegraphics[width=\textwidth]{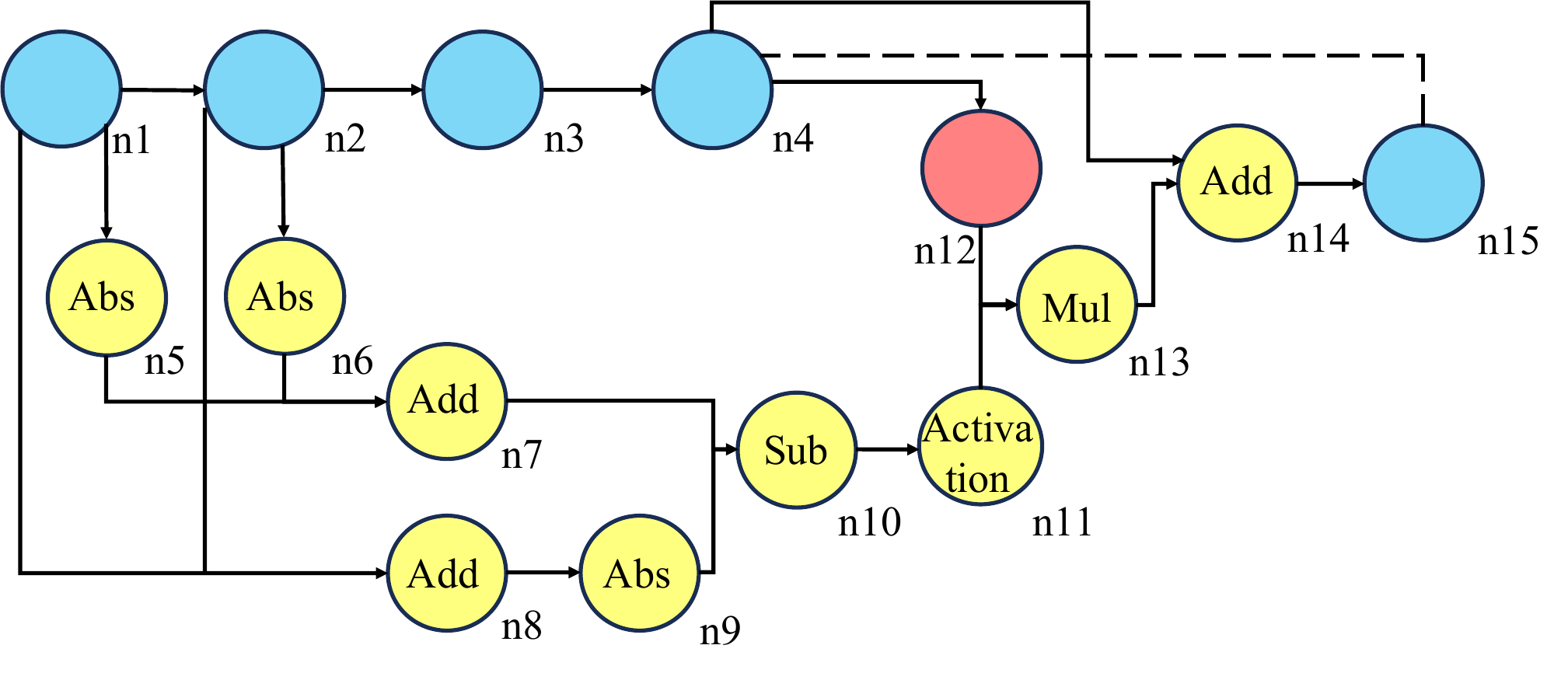}
        \caption{SMR2}
        \label{fig:smr2}
    \end{subfigure}

     \begin{subfigure}[b]{0.49\textwidth}
        \includegraphics[width=\textwidth]{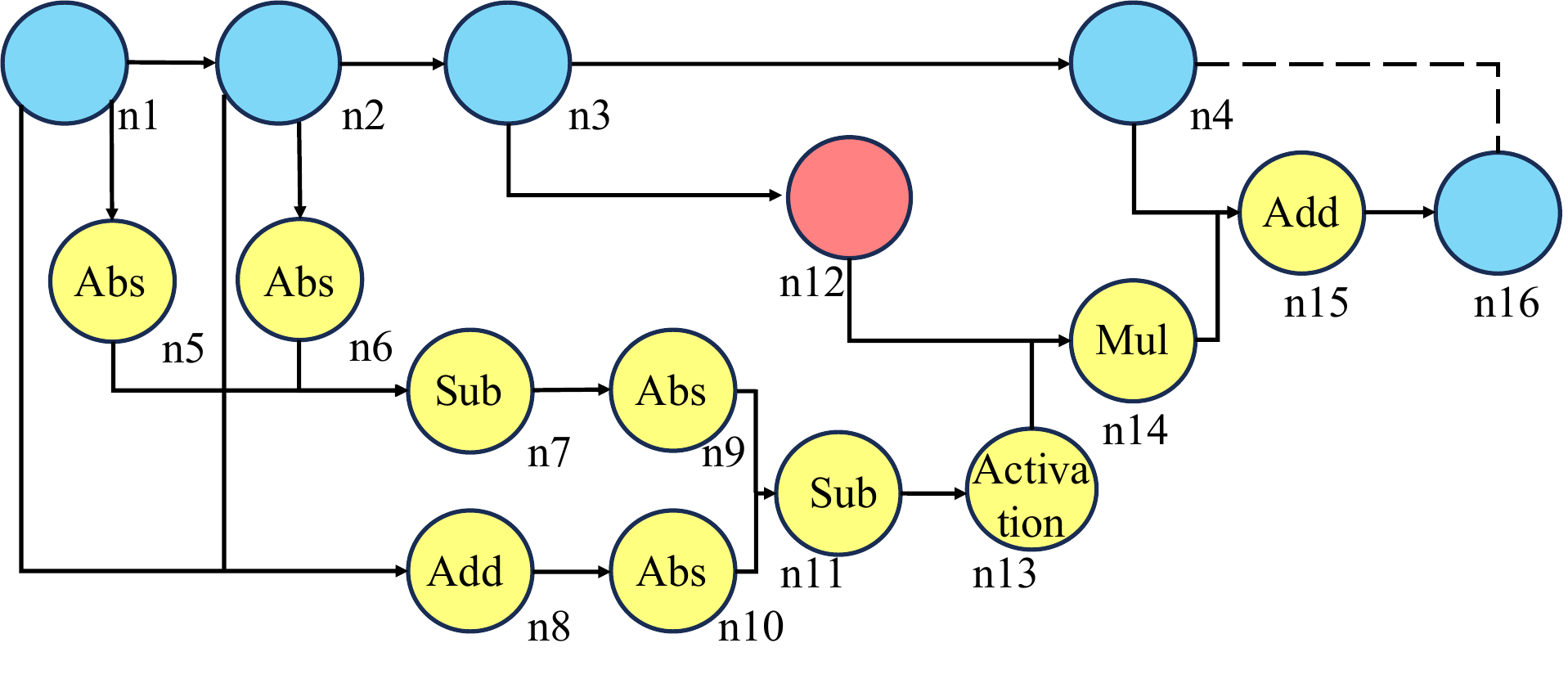}
        \caption{SMR3}
        \label{fig:smr3}
    \end{subfigure}
    \begin{subfigure}[b]{0.4\textwidth}
        \includegraphics[width=\textwidth]{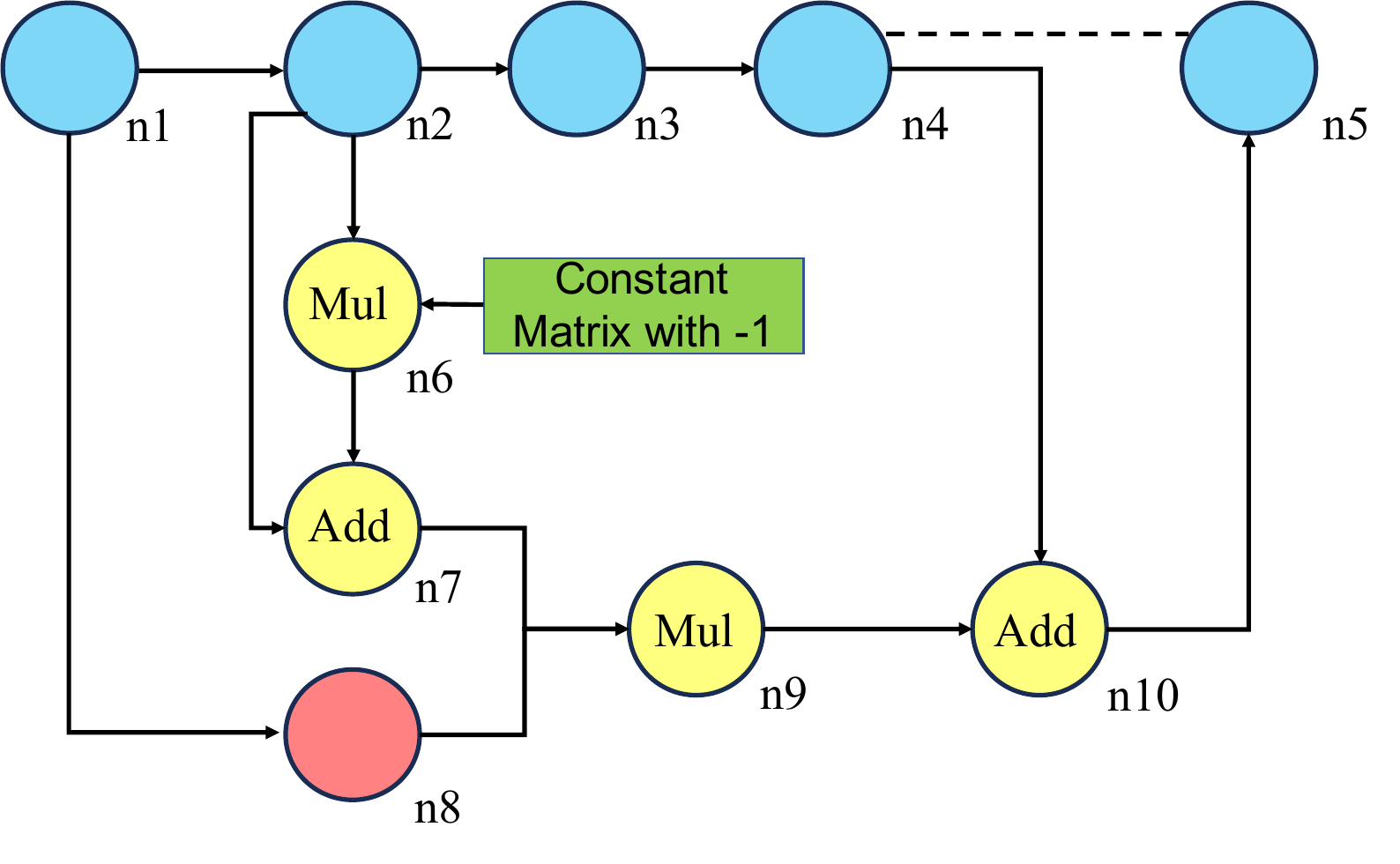}
        \caption{SMR4}
        \label{fig:smr4}
    \end{subfigure}
    
    \vspace{-4mm}
   \caption{Simplified View of Structure MRs}
   
  \parbox{\textwidth}{\footnotesize * The blue nodes represent the original layers of the seed model, the red nodes represent the externally inserted structures, the yellow nodes represent the structure MR, and the green nodes represent the tensors with constant values.  Besides, the dashed lines represent the original connection between layers, while the connection is removed after transformation.}
   \vspace{-4mm}
    \label{fig:smr}
\end{figure}

\textbf{Structure MRs.} To promote the structural diversity of the test inputs and detect more diverse bugs related to different interface combinations,
we design four SMRs to insert external structures consisting of single interfaces or groups of interfaces into the seed model. Based on the SMRs, the generated model with inserted structures must produce the same output as the original seed model.
We present four examples to show how to generate new models based on the SMRs in Fig.~\ref{fig:smr}. We record the output of the node $n_{x}$ in Fig.~\ref{fig:smr} as $O_{n_{x}}$.\looseness=-1

\textbf{SMR1: Complete Square Transformation.} 
Fig.~\ref{fig:smr1} shows how \tool implements the cascade structure of the complete square equation (yellow and green nodes) to ensure output consistency between the new model and the original one after adding external structures (red nodes).
The motivation for this MR comes from the following properties:
(1) Given two constants $a$ and $b$, the result of $(a+b)^2$ is always bigger than zero.
(2) Some specific activation layers (e.g., the ``ReLU'' activation) output zero when the input is non-positive values.
Therefore, we combine these two properties and design SMR1. 
As shown in Fig.~\ref{fig:smr1}, \tool performs a complete square on $O_{n_2}$ and $O_{n_4}$ except for the original connection and multiplies the result by -1 as shown in Fig.~\ref{fig:smr1}. Please note that if the shape and dimension of $O_{n_2}$ and $O_{n_4}$ are different, \tool pads the small one with ``zero'' value to align them.
Since a squared value is always non-negative (i.e., $(O_{n_2} + O_{n_4})^2 \geq 0$), multiplying by -1 guarantees a non-positive result. \tool then inputs the result into an activation layer $n_{11}$ and the outputs are all zero. Meanwhile, \tool inserts external structure $n_{13}$ (e.g., one single DL operator or cascade structure) into the seed model. Then, \tool multiplies the output of inserted layers with zero output (i.e., $O_{n_{11}}$) and adds it to the original layer $n_5$ of the seed model. The output of the transformed model should be consistent with the original one. \looseness=-1

\textbf{SMR2: Absolute Inequality Transformation.} 
Fig.~\ref{fig:smr2} shows how \tool implements SMR2 related to the absolute inequality (yellow nodes) to ensure output consistency between the new model and the original one after adding external structures (red nodes).
The motivation for this MR comes from the following property. Given two constants $a$ and $b$, the inequality $|a|+|b|\geq |a+b|$ always holds true. Therefore,  the opposite result of $|a|+|b|-|a+b|$ is always less than zero. 
Like SMR1, we combine such inequality with the activation layers that convert non-positive input to 0 and design SMR2.
As shown in Fig.~\ref{fig:smr2}, \tool first adds the absolute calculation after $n_1$ and $n_2$ and then adds their result, i.e., adds $O_{n_5}$ and $O_{n_6}$ to layer $n_7$. Besides, \tool calculates the sum of $O_{n_1}$ and $O_{n_2}$ (i.e., $O_{n_8}$). Then \tool subs $O_{n_9}$ with $O_{n_7}$ and inputs the result to the activation layer $n_{11}$.
Based on the absolute inequality, 
i.e., $ | O_{n_1} | + | O_{n_2} | \geq  | O_{n_1}  +  O_{n_2} |$, $O_{n_{10}}$ is always less than or equal to zero.  Therefore, $O_{n_{11}}$ is zero.
Finally, \tool multiplies $O_{n_{11}}$ with $O_{n_{12}}$ to eliminate the effects on the original model calculation. The output of the transformed model should be consistent with that before the transformation. \looseness=-1

\textbf{SMR3: Triangle Inequality Transformation.} 
Fig.~\ref{fig:smr3} shows how \tool implements the cascade structure of the triangle inequality (yellow nodes) to ensure output consistency between the new model and the original one after adding external structures (red nodes).
The motivation for the MRs comes from the following property.
Given two constants $a$ and $b$, the inequality $||a|-|b|| \leq |a+b|$ always holds true. Therefore,  the result of $||a|-|b||-|a+b|$ is always smaller than zero. 
Like SMR1, we combine such inequality with the activation layers and design SMR3.
The details of SMR3 are shown in Fig.~\ref{fig:smr3}. 
\tool first subs the absolute output of  $n_1$ and $n_2$, then achieves its absolute value, i.e., $O_{n_9}$.
Besides, \tool adds $O_{n_1}$ and $O_{n_2}$ and achieves its absolute value, i.e., $O_{n_{10}}$.  Then, \tool subs $O_{n_9}$ with $O_{n_{10}}$ and inputs the result (i.e., $O_{n_{11}}$) to the activation layer $n_{13}$. 
Based on the triangle inequality, 
i.e., $ | |O_{n_1}| - |O_{n_2}| | \leq  | O_{n_1} +  O_{n_2} |$,
$O_{n_{11}}$ is always less than or equal to zero and $O_{n_{13}}$ is zero.
\tool further multiplies $O_{n_{13}}$ with the output of the inserted structure $n_{12}$ to eliminate the effects on the original model calculation. The output of the transformed model should be consistent with that before the transformation.

\textbf{SMR4: Inverse Number Transformation.} 
Fig.~\ref{fig:smr4} shows how \tool implements SMR4 related to ``additive inverse'' (yellow and green nodes) to ensure output consistency between the new model and the original one after adding external structures (red nodes).
The motivation for the MRs comes from the following property.
Given a constant $a$, its opposite number is $-a$, and the result of $a+(-a)$ is also zero. 
Based on this, we design SMR4, and the details are shown in Fig.~\ref{fig:smr4}.
Given one layer $n_2$ of the seed model, \tool first multiplies $O_{n_2}$ with -1 and then adds the result (i.e., $O_{n_6}$) with $O_{n_2}$. Then, \tool multiplies the result with the output of the inserted structures to keep the output consistent with the original model.
\looseness=-1

\textbf{Interface MRs.} 
In addition to SMRs, we migrate two kinds of IMRs focused on (1) input data and (2) parameter settings of single interfaces from existing work~\cite{wang2022eagle,2022metamorphic}, to enhance test input diversity after applying SMRs. Details of the IMRs are as follows.\looseness=-1

\textbf{IMR1: Data Transformation.} These MRs involve equivalent value changes to input data of interfaces, including eight MRs from Chen et al.~\cite{2022metamorphic} related to ``Conv'', ``Batchnorm'', ``Tanh'', ``ReLU'', etc. More specifically, these MRs add some minimum constants or transpose the original input, etc, to ensure consistency between the new and the original output. 

\textbf{IMR2: Parameter Transformation.} Such kind of MRs focuses on changing the values of interface parameters to generate specific outputs, including ten MRs, where six are from Chen et.al~\cite{2022metamorphic} and four MRs are from Wang et.al~\cite{wang2022eagle}. For example, one IMR changes the value of parameters like ``padding'' of ``Conv2d'' to ensure the output consistency between the original interface and the changed interface under the same input.

\begin{wrapfigure}{l}{0.5\textwidth} 
    \includegraphics[width=1.05\linewidth]{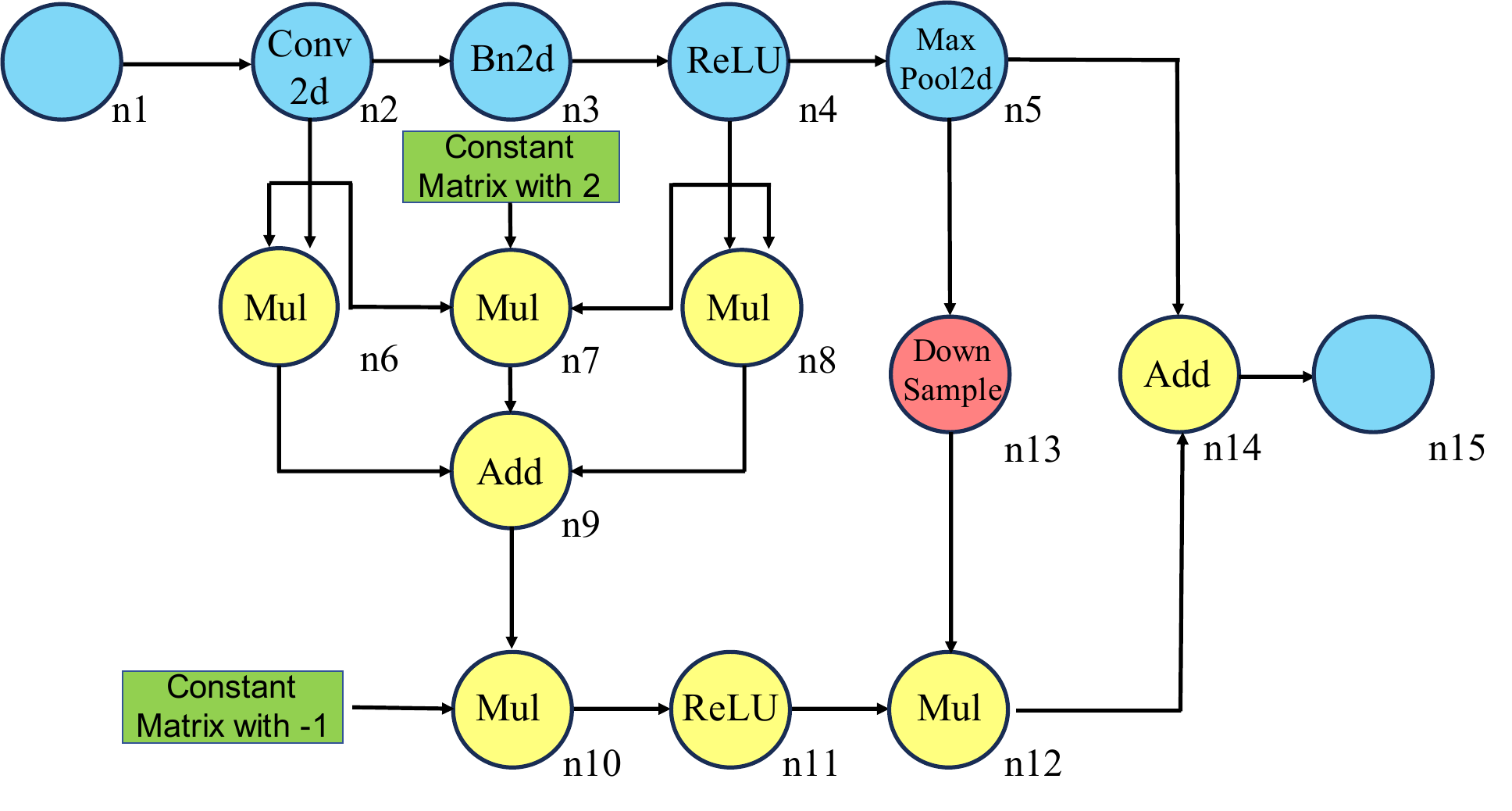}
    \vspace{-6mm}
     \caption{Motivating Example of \tool} 
     \vspace{-4mm}
     \label{fig:mtexample}
\end{wrapfigure}

\textbf{Motivating Example.} To illustrate how \tool applies designed MRs to generate new models and detect bugs, we use a motivating example~\cite{bugcase2} of an accuracy bug (shown in Fig.~\ref{fig:mtexample}) detected by \tool while existing methods cannot. 
The blue nodes represent the layers of the original model, the yellow and green nodes represent the layers to implement our MR, and the red nodes represent the inserted external structure (i.e., one ``DownSample''~\cite{he2016deep} cascade structure). As shown in Fig.~\ref{fig:mtexample}, SMR1 implements the equivalent transformation based on the complete square calculation. The output of the new model must remain consistent with that of the original one based on SMR1.
Model layers $n_6$, $n_7$, $n_8$, and $n_9$ perform complete square calculation on the outputs of nodes $n_2$, $n_3$, and $n_4$. According to the mathematical properties of complete square calculation, the output value of node $n_9$ should be non-negative, and its
negation result is always non-positive (the output
of $n_{10}$). After the ``ReLU'' calculation ($n_{11}$), any non-positive number becomes 0 (the output of $n_{11}$).
Then the output of $n_{11}$ multiplies with the output of the inserted external structure ($n_{13}$) and achieves 0. 
Since adding any number to 0 will not change the original value, the output of $n_{14}$ must equal the output of $n_5$.
However, we find that the generated model violates the constraint of the MRs while the max absolute value of the output of $n_{12}$ is -1048576. 
The feedback from developers shows that
the output accuracy of the ``Mul'' interface of MindSpore cannot correctly express the output
value, which causes an accuracy error. \tool can effectively induce diverse interface groups to trigger such bugs. However, existing metamorphic testing-based methods focus on testing different features of single interfaces and cannot detect such bugs exposed by different interfaces.\looseness=-1

\vspace{-3mm}
\subsection{Model Generation}
\label{sec:seedselection}

This section explains how to select MRs, external structures, and seed models to generate new models. \tool selects SMRs and IMRs and applies them to the seed model, which is enhanced with external structures. If it crashes, it selects a new model from the previously generated models for the next iteration. The above process can be modeled through the MDP definition as introduced in Section~\ref{sec:rllearning}. \looseness=-1

\textbf{Definition 1. (State).} One state $s$ in the model generation process represents one DL model, and the whole test input space is the set of all generated models ${D=\{D_1, D_2, ..., D_n\}}$.\looseness=-1

\textbf{Definition 2 (Action).} Since one ``state'' is a DL model, an action $a$ is the transformation of the MRs, including the SMRs we designed and the IMRs from previous studies. The MRs insert new layers, change the parameter settings, and equivalently modify the value of input data of the existing model to generate new models, i.e., transform one ``state'' to a new ``state''. \looseness=-1

\textbf{Definition 3 (Policy).} A policy $\pi$ decides how to select the MRs. We adopt the QR-DQN strategy with the $\epsilon$-greedy strategy as the policy $\pi$. The QR-DQN strategy efficiently captures valuable ``states'', i.e., those models that can trigger bugs across the entire test input space. Based on this, \tool can adapt to the vast and unstable test input space of extensive DL models with complex structures, diverse interface groups, varying parameter settings, and large-scale weights. Besides, the $\epsilon$-greedy strategy introduces randomness to avoid local optima by dynamically adjusting epsilon, accelerating convergence, and enhancing the bug detection performance of \toolnospace. Besides, we take the random strategy for selecting IMRs.
\looseness=-1

\textbf{Definition 4 (Reward). } \tool focuses on exploring, i.e., more diverse interface combinations, parameter settings, and data shapes, of the ``states'' and calculates the arithmetic average of three diversity metrics used in the previous studies~\cite{li2023comet} as the ``reward'' function.\looseness=-1

\myz{As defined above, \tool selects one seed model, SMR, IMR, and external structures in each round, requiring multiple choices. These choices serve different purposes, e.g., selecting MRs enhances model diversity while selecting models identifies valuable historical models for further testing. Therefore, \tool employs different strategies for each factor, rather than using one single strategy, which may limit the diversity of generated test inputs (e.g., one strategy that effectively selects historical models might repeatedly choose the same MRs).
We comprehensively adopt three different strategies, i.e., combining the QR-DQN~\cite{dabney2018distributional} and $\epsilon$-greedy strategy~\cite{mnih2015human} for selecting SMRs and adopting the Thompson Sampling strategy~\cite{thompson1933likelihood} for selecting seed models, respectively. 
Specifically, the seed model is input into the neural network of QR-DQN, which calculates the mean quantile for all SMRs. The SMR with the highest value is selected and applied to the seed model. The generated model is then evaluated by calculating its reward and updating its value.
By further combining with the $\epsilon$-greedy strategy, \tool can introduce randomness and avoid local optima (i.e., always selecting the MR with the maximum reward but ignoring exploring other buggy branches). Furthermore, the Thompson Sampling strategy can accurately prioritize unexplored valuable generated models by dynamically updating the reward distribution based on the posterior probability distributions (e.g., beta distribution~\cite{kotz2004beyond} in our study), further improving the stability and efficiency of \toolnospace.
}

The details of the whole model generation process are shown in Algorithm~\ref{alg:modelMeta}.
\tool first initializes the parameters (Line $1$), including 
(1) the seed model $d$,
(2) the experience pool $ReplayPool$ that stores the history generation experience, i.e., the generated model and relevant reward in one iteration,
(3) the quantum Q-network $QunantumQ$ and its copy $TargetQ$ which is used to guide the selection of MRs, and (4) the list $BugList$ stores all the detected bugs.
Then it executes $N$ rounds (Line $2$) to generate models based on the MRs iteratively.
\tool utilizes the prediction of $QunantumQ$ combined with the $\epsilon$-greedy strategy to select the SMR $MR_{Structure}$: it generates a probability $p$ (Line $3$) from the symmetric distribution (average is 0, and the standard deviation is 1) and compares it with the preset threshold $\epsilon$ (Line $4$). If $p$ is smaller than $\epsilon$, \tool randomly selects the SMR $MR_{Structure}$ (Line $5$). Otherwise, it selects the SMR $MR_{Structure}$ based on the prediction of $QuantumQ$ (Line $7$). Meanwhile, \tool randomly selects one candidate layer $InsertLayer$ for $MR_{Structure}$ (Line $8$). and one IMR $MR_{Interface}$ to promote the diversity of generated new model (Line $9$). Then \tool first applies $MR_{Structure}$ and insert $InsertLayer$  to $d$. Then $MR_{Interface}$ is applied on $d$ by randomly selecting one layer to generate $d'$ (Line $10$). 

If it crashes (Line $11$), \tool assigns the reward with -1 to penalize the selection of $MR_{Structure}$ and $MR_{Interface}$ in the current iteration (Line $12$), then select a new seed model from the experience pool $ReplayPool$ using the Thompson Sampling strategy~\cite{thompson1933likelihood} (Line $13$), which is effective for exploration in uncertain spaces. It assigns the \textit{Beta} distribution~\cite{johnson1995continuous} to conduct sampling for each non-crash generated model and then update its parameters.
If the generated $d'$ does not crash (Line $14$), \tool calculates the average value of three diversity metrics (as introduced in Section~\ref{sec:measurements}) as the reward $r$ and records it as a non-terminating state, i.e., $Done$ is set to ``False'' (Line $15$). Besides, all the key parameters, including the reward $r$, generated model $d'$, are recorded in the pool $ReplayPool$ (Line $16$). \tool updates $QuantumQ$ by minimizing the difference between its prediction of $QuantumQvalue$ and the reward value $TargetQvalue$ predicted by $TargetQ$ about $d'$ (Lines $17$-$19$). It also updates $TargetQ$ by copying the weight of $QuantumQ$ (Line $20$). Besides, \tool dynamically reduces $\epsilon$ to converge the model generation (Line $21$). At the end of each iteration, \tool adopts $d'$ as the seed model for the next round (Line $22$).

\vspace{-2mm}
\subsection{Bug Detection}
\label{sec:bugdetection}
Detecting diverse kinds of bugs is critical to the effectiveness of the testing methods.
However, existing methods~\cite{wang2022eagle,ding2017validating,chen2024miss} only analyze if the original and generated test inputs achieve consistent execution results, ignoring bugs that may appear in intermediate execution states, like resource usage and execution time, or in combinations of multiple interfaces.
To overcome this limitation, 
\tool detects bugs based on all the generated models by analyzing resource (i.e., memory and GPU) usage, execution time, model outputs, training loss, gradients, and crashes (Lines $23$-$26$). Finally, \tool returns the detected bugs $BugList$ (Line $27$).
Please note that the thresholds related to detecting bugs are different for different kinds of bugs and models with different scene tasks (e.g., image classification). Therefore, we run all the seed models, collect preliminary results, and analyze the typical cases to determine the specific values for thresholds. The specific values for different bugs on each model can be found on our website~\cite{sharelink}. Test oracles for these bugs are as follows. \looseness=-1

\textbf{Resource Bugs.} We monitor the memory usage of each model in $LeftModels$ across CPU and GPU modes. Particularly, we calculate the ratio of memory and GPU usage size of each model and detect whether the ratio is larger than others or fails to release resources after execution, which means the model uses unreasonable resources.

\textbf{Efficiency Bugs.} We record the execution time of each generated model in $LeftModels$ on CPU and GPU modes. Specifically, we calculate the time ratio each model spends on CPU and GPU and check if it exceeds that of other models, indicating abnormal computing behavior.

\textbf{Crash Bugs.} Models meet crash when applying MRs or calculating input data, excluding false positives caused by unreasonable executions.

\textbf{Accuracy Bugs.} We record the (1) forward outputs, (2) training loss, and (3) gradients of each generated model, and calculate the differences between them. Specifically, we adopt the Chebyshev distance~\cite{deza2009encyclopedia} as the difference metric for model outputs. The definition is as follows. \looseness=-1

\begin{equation}
    D^{M,N}_{f_{L_i}} (x) = Max (|M_{f_{L_i}} (x) - N_{f_{L_i}} (x) |)
    \label{equ:layerdis} 
\end{equation}

Given a DL model with $n$ layers $ f= \langle L_0, L_1, \cdots, L_n \rangle$ and an input tensor $x$, the output of the $i$-th layer is recorded as $ f_{L_i} (x)$. $D^{M,N}_{f_{L_i}} (x)$ measures the output distance of the model $M$ and the model $M'$ on layer $i$. An accuracy bug is detected when the value of $D^{M,N}_{f_{L_i}} (x)$ exceeds the preset thresholds. For the loss and gradient, we follow the settings of Gu et al.~\cite{gu2022muffin} to detect the accuracy bugs in them. Besides, if inconsistent outliers exist (e.g., NaN or Inf) across different generated models, we consider it to expose an accuracy bug.

\section{Experiment Design}

\subsection{Research Questions}
\label{sec:rqs}
In this study, we aim to investigate the following research questions:

\(\bullet\)
\textbf{RQ1: How does \myz{\tool} perform when compared with the SOTA baselines?}
We compare \tool with selected baseline methods from the diversity generated test inputs, test coverage, detected bugs, and execution time to evaluate its effectiveness. \looseness=-1

\(\bullet\)
\textbf{RQ2: What types of bugs does \myz{\tool} detect?}
We investigate the root causes, symptoms, and trigger conditions of the bugs detected by \tool to further analyze the bug detection ability of \toolnospace. \looseness=-1

\(\bullet\)
\textbf{RQ3: How do different MRs contribute to \myz{\tool}?}
We conduct an ablation experiment to investigate the effects of different structure MRs on \tool based on the generated test inputs. \looseness=-1

\(\bullet\)
\textbf{RQ4: How does the QR-DQN strategy contribute to \myz{\tool}?}
We conduct an ablation experiment to investigate whether the QR-DQN strategy can effectively and efficiently guide the model generation process based on the generated test inputs.

\subsection{Benchmark}
\label{sec:benchmarks}
\textbf{DL Frameworks.} 
We select three popular DL frameworks (i.e., MindSpore 2.2.0~\cite{mindspore}, PyTorch \myz{2.3.0}~\cite{torch}, and ONNX 1.14.0~\cite{onnx}). They are widely adopted under testing in previous DL framework testing studies~\cite{wang2020lemon,2020audee,li2023comet}. Specifically, PyTorch is a well-known DL framework that has outstanding characteristics like dynamic computation graphs and user-friendly design, enabling rapid prototyping and seamless platform deployment. MindSpore is an innovative DL framework tailored for diverse environments, with an efficient execution engine that optimizes resource management and task execution. ONNX is an open-source framework offering interoperability between DL models, enabling stable integration and deployment across different platforms.

\textbf{DL Models and Datasets.} 
To evaluate \tool comprehensively, we consider 17 DL models with ten kinds of tasks, ranging from image classification and object detection to scene recognition, as shown in Table~\ref{tab:modelstatistics}.
Specifically, for image classification, we choose four common DL models (i.e., VGG16, ResNet50, Mobilenetv2, and Vision Transformer (VIT)); 
For object detection, we choose four models (i.e., \myz{Yolov3-DarkNet53, Yolov4, SSD-ResNet50-FPN, SSD-Mobilenetv1}); 
For semantic segmentation, we choose two models (i.e., Unet and Deeplabv3).
We also chose one text classification model (TextCNN), one anomaly detection model (Patchcore), one defect detection model (SSIM-AE), one key point detection model (OpenPose), and one scene recognition model (CRNN).
Furthermore, we add one image generation model (i.e., SRGAN~\cite {ledig2017photo}) and one text generation model (i.e., GPT-2~\cite{radford2019language}) into our benchmark since the generation-based DL models have become popular in real industry applications. Please note that the depth of GPT-2 is adjustable, and we set its depth based on the resource constraints of our execution environment. \looseness=-1

\begin{table}[!ht]
  \centering
  \caption{Statistics of Original DL models}
    \begin{tabular}{ccccc}
    \toprule
    \textbf{Scene Task} & \textbf{Model} & \textbf{Depth} & \textbf{Width} & \multicolumn{1}{c}{\textbf{Parameters}} \\
    \midrule
    \multirow{4}[1]{*}{Image classification} & VGG16 & 53    & 4096  & 134,314,186 \\
          & ResNet50 & 126   & 2048  & 23,581,642 \\
          & MobileNetv2 & 141   & 1280  & 2,270,794 \\
          & VIT   & 150   & 3072  & 87,423,754 \\ \hline
    \multirow{4}[1]{*}{Object Detection} & SSD-ResNet50-FPN & 232   & 2048  & 33,370,814 \\
          & SSD-MobilenetV1 & 143   & 1024  & 11,329,461 \\
          & Yolov3-Darknet53 & 222   & 1024  & 62,001,757 \\
          & Yolov4 & 334   & 1024  & 65,741,916 \\
    \midrule
    \multirow{2}[2]{*}{Semantic Segmentation} & Deeplabv3 & 263   & 2048  & 58,149,077 \\
          & Unet  & 53    & 1024  & 31,029,698 \\
    \midrule
    Text Classification & TextCNN & 12    & 96    & 859,146 \\
    \midrule
    Anomaly Detection & PatchCore & 129   & 2048  & 68,951,464 \\
    \midrule
    Defect Detection & SSIM-AE & 39    & 500   & 2,720,768 \\
    \midrule
    Key Point Detection & OpenPose & 112   & 512   & 52,311,446 \\
    \midrule
    Scene Recognition & CRNN  & 29    & 512   & 4,339,621 \\
    \midrule
    Text Generation & GPT-2 & 112   & 768   & 175,620,108 \\
    \midrule
    Image Generation & SRGAN & 117   & 1024  & 1,554,691 \\
    \bottomrule
    \end{tabular}%
  \label{tab:modelstatistics}%
\end{table}%

\subsection{Baseline Methods}
\label{sec:baselines}
We adopt two state-of-the-art DL framework testing methods based on metamorphic testing as baselines, i.e., EAGLE~\cite{wang2022eagle} and Meta~\cite{chen2024miss}. We also select two popular model-level differential-based methods, i.e., COMET~\cite{li2023comet} and Muffin~\cite{gu2022muffin} as baselines.
Specifically, EAGLE designs 16 equivalence rules based on optimization, data formats, and specific interface characteristics to detect bugs by checking the inconsistencies between original and equivalent test inputs.
Meta is the metamorphic testing-based method that designs 18 MRs based on 10 interfaces' input data and parameters to effectively explore the high-dimensional input tensors and the discrete parameter settings.  COMET employs a variety of mutation operators, including weight, input, parameter, and structure mutations guided by the MCMC strategy to generate more new models with diverse structures. Muffin generates models by using two preset model structure templates. It detects bugs during training stages, including forward computation, loss calculation, and backpropagation.\looseness=-1

\subsection{Measurements}
\label{sec:measurements}
We evaluate \toolnospace's performance based on the diversity of generated test inputs, test coverage, bug detection, and test efficiency. 

\textbf{Measurements for the Diversity of Generated Test Inputs.} Specifically, we adopt three diversity metrics that are widely adopted in previous studies~\cite{li2023comet} to evaluate the generated models.\looseness=-1

\(\bullet\)
\textit{Layer Input Coverage (LIC).} It measures the diversity of the input data of one layer from the perspective of the input shape, dimension, and data types. 
The input information of the middle layer includes shape, dimension, and data type. 
Supposing the total number of data types is $N_{type}$, 
the total number of dimensions is $N_{dim}$, 
and the total number of shapes is $N_{shape}$. 
The number of data types covered is $n_{type}$, 
the number of dimensions covered is $n_{dim}$, 
and the number of shapes covered is $n_{shape}$. 
The layer input coverage $LIC$ is calculated as the result of the sum of $N_{type}$, $N_{dim}$, and $N_{shape}$, divided by the sum of $n_{type}$, $n_{dim}$, and $n_{shape}$, i.e., 
$LIC = \frac{n_{type}+n_{dim}+n_{shape}}{N_{type}+N_{dim}+N_{shape}}$.

\(\bullet\)
\textit{Layer Parameter Coverage (LPC).} It measures the diversity of parameter settings, i.e., the layers with different values of parameters. Supposing the middle layers $l$ and $l'$, and only when their internal parameters (e.g., ``in/out channels'', ``kernel\_size'', and ``stride'' of Conv2d) are the same, $l$ is considered equivalent to $l'$. Therefore, the number of different middle layers in the current models is $n_{ut}$, and the number of different edges in all models is $N_{ut}$. Therefore, the $LPC$ is calculated as the result of  $N_{ut}$ divided by $n_{ut}$, i.e., $LPC = \frac{n_{ut}}{N_{ut}}$. \looseness=-1

\(\bullet\)
\textit{Layer Sequence Coverage (LSC).} It measures the diversity of the layer sequence, i.e., the combinations of different layers in models. In the definition of $LSC$, the layer sequence refers to groups of two different framework interfaces. Supposing the total number of all the interface groups that appeared in the execution is $AP_{sum}$ and the number of the interface groups of the current generated models is $AP_{cur}$. The layer sequence coverage $LSC$ is calculated as the result of $AP_{sum}$ divided by $AP_{cur}$, i.e., $LSC = \frac{AP_{cur}}{AP_{sum}}$. \looseness=-1

\myz{\textbf{Measurements for Test Coverage.} We use API and code coverage to assess whether \tool can thoroughly cover testing the implementations of DL frameworks. API coverage is essential since bugs are triggered only when the corresponding interfaces are covered, measured using the ``Coverage.py'' tool. Code coverage, a standard metric in traditional software testing~\cite{brosgol2011178c,legunsen2016extensive,zhang2018hybrid}, focuses on tracing C/C++ code using LCOV~\cite{lcov} and GCOV~\cite{gcov}. This emphasis on C/C++ is due to its foundational role in operator implementation, as most high-level Python interfaces eventually invoke underlying C/C++ code. }\looseness=-1

\textbf{Measurements for Bug Detection.} We first analyze the detected bugs of \toolnospace, i.e., the number and types of bugs. Then, we compare with existing methods from the following perspectives: (1) the confirmed and fixed bugs and (2) the bugs \tool can detect while baseline methods cannot. 
\myz{Three authors involved in implementing \tool and baseline methods manually review all reported bugs to verify their status (e.g., confirmed or fixed) and filter out false positives. 
For bugs identified in prior studies, we first validate open bug links and analyze recent developer comments to update their status. Each author independently assesses comments, developer feedback, and root causes. If the authors cannot agree on the bug state, they will discuss it and make a final decision. A bug is marked as ``rejected'' if all three authors agree that the developers' comments indicate rejection. Confirmed bugs with fixes or PR links are labeled fixed; others are categorized as ``waiting for fix''. 
For bugs detected by existing methods or \tool on our benchmark, the authors reproduce the cases, analyze symptoms, and exclude false positives. The remaining cases are compiled with code snippets and symptom descriptions and then submitted to framework communities for developer review. After receiving developer feedback, authors verify the bug status based on report comments and decide the final bug state when they reach a consensus. Bugs with confirmed or rejected comments from developers checked by the three authors are marked as ``confirmed'' or ``rejected''. Bugs with unclear comments are marked as ``waiting for confirmation''. Confirmed bugs with explicit fix responses checked by the three authors are marked as ``fixed''; others are marked as ``waiting for fix''.
}
\looseness=-1

\textbf{Measurements for Test Efficiency.} We count the average execution time of each method to compare the test efficiency of each method, ranging from test input generation to bug detection. \looseness=-1

\subsection{Implementations}
\label{sec:implementations}
\myz{To fairly evaluate the performance of \tool and baseline methods, we implement them on the same version of MindSpore 2.2.0, PyTorch 2.3.0, and ONNX 1.14.0. Specifically, we reimplement the original COMET and Muffin on PyTorch and MindSpore based on their original tools and convert the generated models to ONNX. Additionally, we reimplement EAGLE for MindSpore and ONNX, and Meta for ONNX. 
All experiments are conducted on the Intel(R) Xeon(R) Gold 6226R CPU @2.90GHz machine with 256GB of RAM, Ubuntu 20.04.4 LTS, and eight 24GB Nvidia GTX 3090. The CUDA
version is 11.1, and the Anaconda and Python versions are 4.14.0 and 3.7, respectively.
}\looseness=-1

The parameters of \tool include execution rounds, QR-DQN settings, and bug detection thresholds. We set the execution round limit to 100, following prior work~\cite{wang2020lemon}, meaning \tool terminates after 100 rounds of model generation. As the rounds increase, more complex models are generated, enhancing test input diversity. However, this also increases the risk of generating invalid models (e.g., the outputs of such models exceed the framework's effective accuracy range). The increased complexity demands more time for both model generation (selecting appropriate seed model layers for applying MRs) and bug detection (executing the generated models). We find that 100 rounds strike the best balance, achieving significant model diversity with fewer invalid models and optimal performance across all metrics. Thus, we set 100 as the execution limit.
For the QR-DQN's settings, like the number of quantiles, we follow the recommendations from previous studies~\cite{dabney2018distributional} to achieve robust and effective performance. For the thresholds of bug detection, we conduct large-scale experiments, manually analyze the results, and conclude an entire list of the specific values for each model and each defect type, which can be found on our repositories~\cite{sharelink}. All internal parameter settings of baseline methods follow the default recommendation in their previous studies~\cite{gu2022muffin,li2023comet,2022metamorphic} and we set the execution round to 100 for all baseline methods.
\looseness=-1

\section{Result Analysis}
\subsection{RQ1: Performance Evaluation}
\textbf{Design.} \myz{We evaluate the performance of \tool and baseline methods from the following aspects: (1) the diversity of generated models, (2) test coverage (API and code coverage), (3) detected bugs, and (4) execution time. }

\begin{table}[htbp]
  \centering

  \caption{\myz{Comparative Results across \tool and Baseline Methods}}

  \resizebox{\linewidth}{!}{
    \begin{tabular}{c|ccc|c|c|cc|c}
    \hline
    \multirow{2}[2]{*}{Method} & \multicolumn{3}{c|}{Model Diversity on Average} & \multirow{2}[2]{*}{Code Coverage (total 124216)} & \multirow{2}[2]{*}{API Coverage (total 1131)} & \multicolumn{2}{c|}{Execution Time (Seconds)} & \multirow{2}[2]{*}{False Positive} \\
          & LSC   & LIC   & LPC   &       &       & Generation & Detection &  \\
    \hline
    EAGLE & -     & -     & -     & 28783 (23.172\%) & 637 (56.322\%) & 1.141 & 1.456 & 36.364\% \\
    Meta  & -     & -     & -     & 23421 (18.855\%) & 589 (52.078\%) & - & 0.011 & 33.333\% \\
    Muffin & 0.001 & 0.060 & 0.018 & 29123 (23.445\%) & 660 (58.355\%) & 14.45 & 1.857 & - \\
    COMET & 0.002 & 0.058 & 0.021 & 31594 (25.435\%) & 759 (67.109\%) & 6.316 & 1.441 & 40.000\% \\
    ModelMeta & 0.002 & 0.062 & 0.023 & 35746 (28.777\%) & 890 (78.691\%) & 5.997 & 11.07 & 27.910\% \\
    \hline
    \end{tabular}%
    }
    
  \label{tab:rq1results}%
\end{table}%

\begin{table}[]
  \centering
  \caption{\myz{Bugs Detected by Different MRs}}
  
    \begin{tabular}{c|c|c|c|c|c|c|c}
    \hline
    MRs   & \multicolumn{1}{l|}{SMR1} & \multicolumn{1}{l|}{SMR2} & \multicolumn{1}{l|}{SMR3} & \multicolumn{1}{l|}{SMR4} & \multicolumn{1}{l|}{IMR} & \multicolumn{1}{l|}{Confirmed} & \multicolumn{1}{l}{Fixed} \\
    \hline
    Resource Bug & 3     & 1     & 1     & 0     & 0     & 4     & 4 \\
    Efficiency Bug & 1     & 0     & 0     & 1     & 0     & 2     & 0 \\
    Crash Bug & 3     & 3     & 4     & 3     & 1     & 12    & 6 \\
    Accuracy Bug & 1     & 2     & 4     & 1     & 2     & 9     & 1 \\
    \hline
    Total & 8     & 6     & 9     & 5     & 3     & 27    & 11 \\
    \hline
    \end{tabular}%
    
  \label{tab:mrbugs}%
\end{table}%

\myz{\textbf{Comparative Study on the Diversity of Generated Test Inputs.} Notice that EAGLE and Meta cannot calculate the value of LSC, LIC, and LPC since their test inputs are single interfaces. Therefore, we only compare them with \tool based on the covered interfaces. 
As shown in the sixth column of Table~\ref{tab:rq1results}, \tool achieves a 51.1\% and 39.7\% improvement over Meta and EAGLE, respectively, in terms of the number of covered interfaces. This is because \tool covers more interfaces related to calculating loss and gradients.
Compared to the two model-level baseline methods, \tool achieves an average improvement of 5.5\% (no improvement on $LSC$, $\uparrow$6.9\% on LIC, and $\uparrow$9.5\% on LPC) and 43.7\% ($\uparrow$100\% on LSC, $\uparrow$3.33\% on LIC, and $\uparrow$27.8\% on LPC) over COMET and Muffin, as shown in the second to fourth columns of Table 2.
}\looseness=-1

\myz{\textbf{Comparative Study on Test Coverage.}} \myz{Please note that given the high cost of collecting coverage across different DL frameworks and the commonality of testing practices and bugs across DL frameworks~\cite{chen2023toward,jia2021unit,nejadgholi2019study,yang2022comprehensive}, we focus on collecting API and line coverage from PyTorch to evaluate each method. As shown in the fifth and sixth columns of Table~\ref{tab:rq1results}, \tool achieves the highest line coverage and API coverage compared with the baseline methods. Specifically, \tool outperforms existing methods with an average 28.2\% improvement in code coverage and an average 35.7\% improvement in API coverage. This is because \tool can cover more interfaces related to the model training process, including calculating loss and gradients, which cannot be covered by existing methods.}

\begin{table}[]
  \centering
  
  \caption{\myz{Bug Detection Results across Different DL Frameworks}}
    \begin{tabular}{c|c|c|c}
    \hline
    Method & PyTorch & MindSpore & ONNX \\
    \hline
    ModelMeta & (10, 2, 3, 7) & (13, 0, 8, 5) & (4, 2, 0, 4) \\
    EAGLE & (2, 2, 0, 2) & (2, 8, 2, 0) & (0, 0, 0, 0) \\
    Meta  & (1, 0, 1, 0) & (1, 2, 1, 0) & (0, 0, 0, 0) \\
    COMET & (2, 2, 1, 1) & (3, 1, 1, 2) & (0, 1, 0, 0) \\
    Muffin & (0, 0, 0, 0) & (0, 0, 0, 0) & (0, 0, 0, 0) \\
    \hline
    \end{tabular}
  \label{tab:rq1bugs}%

\end{table}

\textbf{Comparative Study on Bug Detection.} Table~\ref{tab:rq1bugs} shows the bug detection results on our benchmark for \tool and other baseline methods (excluding bugs detected in previous studies). Each cell of Table~\ref{tab:rq1bugs} presents a quadruple: (confirmed bugs, bugs awaiting confirmation, fixed bugs, and bugs awaiting fix).
COMET reports two NaN bugs and two crash bugs in PyTorch. Of these, two NaN bugs related to the ``ReLU'' and ``Depthwise Conv2d'' interfaces and the two crash bugs related to parameter implementation errors are waiting for confirmation. In MindSpore, two crash bugs (unsupported data type) and one accuracy bug (inconsistent output across CPU and GPU) are confirmed. Besides, COMET reports four crash bug cases in ONNX, all stemming from conversion failures of models generated by MindSpore or PyTorch. Three of them have been excluded by manual inspection, and the remaining one, related to incomplete ONNX functionality, is awaiting confirmation. Muffin reports no bugs in our benchmarks. 
EAGLE identifies four confirmed output outlier bugs across CPU and GPU, with ten additional bugs involving inconsistencies between generated and original inputs. 
Meta detects two bugs in the ``torch.nn.functional.max\_pool2d'' interface and ``mindspore.ops.tanh'' interface when switching data formats from ``NHWC'' to ``NCHW'', which are both confirmed. Two crash bugs related to MindSpore’s unsupported CPU data format are awaiting confirmation. \tool reports 43 error cases, 39 of which are submitted for developer confirmation (four excluded after manual review). Among them, eight are rejected, 27 confirmed, and 11 fixed. Specifically, for the seven resource bug cases, two are false positives (one due to a non-reproducible memory leak and another from environment setup errors), with four confirmed and one awaiting confirmation. Of the three efficiency bugs, one is a false positive (its time cost is deemed acceptable by developers), with two confirmed. Six of the 20 crash bugs are false positives (four due to wrong environment setup and two from framework safety checks), with 12 confirmed and two awaiting confirmation. Of the 13 accuracy bugs, three are false positives (related to abnormal values), with nine confirmed and one awaiting confirmation. Besides, the false positive results for other baseline methods are shown in the final column of Table~\ref{tab:rq1results}: EAGLE (36.36\%, reporting 22 bugs, with eight rejected), Meta (33.3\%, reporting six bugs, with two rejected), and COMET (40\%, reporting 15 bugs, with six rejected).

We also analyze the bugs detected by baseline methods as reported in their previous studies. Meta identifies 41 bugs, including unsupported mode errors, input validation errors, and precision errors. EAGLE detects 25 bugs; after excluding one false positive, 24 remain, comprising 19 inconsistency bugs (e.g., differing outputs under different precision settings) and five crash bugs. Muffin reports 39 bugs after excluding five false positives, including 18 inconsistency bugs (12 in forward outputs, two in loss computation, three in backward computation, and one NaN bug) and 21 crash bugs. However, detailed information is unavailable on their website. COMET identifies 32 bugs, covering conversion tool bugs, wrong output bugs, NaN bugs, resource bugs (e.g., core dumps from DL operators), and interface implementation bugs. 

Compared with existing methods, \tool can detect crash bugs and output accuracy bugs that arise from complex interface interactions, as identified by tools like COMET and Muffin, while reducing false positives caused by framework conversion thanks to our designed MRs. Besides, \tool is also effective at detecting accuracy and single-interface implementation bugs like Meta and EAGLE, due to the incorporation of interface-level MRs that enhance test input diversity.

\myz{\textbf{Comparative Study on Execution Time.}} \myz{\tool takes an average of 17.1 seconds, compared to EAGLE (2.6s), Meta (0.01s), COMET (7.7s), and Muffin (16.3s), as shown in the seventh and eighth columns of Table~\ref{tab:rq1results}. \tool takes longer than existing approaches due to additional bug detection steps, including monitoring memory usage, execution time, and analyzing loss and gradient calculations. However, the extra time is worthwhile because \tool detects new bug types that existing baselines cannot.}\looseness=-1

\finding{1}{
 \textbf{Answer to RQ1.} 
    Though \tool requires more time, it provides higher test coverage by introducing model training interfaces related to loss and gradient calculations. It can detect new bugs, such as resource and efficiency bugs, that existing methods cannot.
}\looseness=-1

\subsection{RQ2: Bug Study}
We present the typical cases under four kinds of bugs as introduced in Section~\ref{sec:bugdetection} and analyze the symptoms, trigger conditions, and root causes. The counts of the detected, confirmed, and fixed bugs with different types and their triggered MRs are shown in \myz{Table}.~\ref{tab:mrbugs}.

\textbf{Resource Bugs.}
Among the five resource bugs, one involves illegal memory access, one is related to memory leaks, and the remaining three involve abnormal behaviors in GPU/memory allocation.
For example, \tool detects a bug~\cite{resourcebugcase} in the ``mindspore.ops.slice'' interface, where it fails to allocate GPU resources. Specifically, during the model generation process for the ``VIT'' model, the fifth generated model logs the exception: ``For SliceGrad, the CUDA kernel fails to run, error number 9.'' This occurs during gradient calculation in back-propagation. 
The bug is traced to the interfaces of ``SMR2'' where outputs are aligned, specifically within the ``mindspore.ops.slice'' interface, which extracts values from the given data dimension. Interestingly, the generated model executes correctly and the bug cannot be reproduced in the CPU environment. Developers have confirmed this bug and plan to update the interface. Existing metamorphic-based methods cannot detect this bug, as the trigger condition is related to incorrect resource allocation during the gradient calculation process of the generated models. \looseness=-1

\textbf{Efficiency Bugs.} 
One of the two efficiency bugs concerns the optimization mechanism of PyTorch, while another is related to the implementation error of MindSpore's single interface, i.e., the ``mindspore.nn.Flatten''.
For example, \tool detects one bug in the interface of ``torch.compile'', which is used to optimize the execution of the PyTorch programs. Specifically, we execute ``VGG16'' for five iterations and invoke ``torch.compile'' (one kind of IMR) to optimize the calculation efficiency of the generated model. However, we find that the new model, after applying the IMRs, executes slower than the original one, i.e., the original model executes 0.04 seconds while the new model executes 11.59 seconds under the same data input. While the IMR triggers this bug, we argue that our SMR4 is crucial for triggering this bug, as the fifth model generated by SMR4 successfully triggers the framework optimization and exposes the bug. Existing metamorphic testing-based methods, which focus on single interfaces, cannot detect this bug or generate test inputs that effectively trigger framework optimization on diverse interface groups. When we report it to developers, they acknowledge that the invoking mechanism of the JIT, i.e., the compiler of PyTorch, needs to improve. Specifically, when executing ``torch.compile'', the JIT needs to further compile other programs among other kernels like oneDNN~\cite{onednn2021} and oneMKL~\cite{onemkl2021}. They have confirmed this bug, and it needs to be fixed.

\textbf{Crash Bugs.} 
Among 14 crash bugs, five bugs are related to the conversion or export failure, three bugs are related to the execution failure of single interfaces on specific environments, and the remaining six bugs are involved in implementation errors or incomplete functionality of the whole framework.
For example, when applying SMR3 with an external structure inserted, i.e., the single interface ``mindspore.ops.ResizeBilinear'', the generated model crashes during input data calculation~\cite{crashbugcase}. The incomplete implementation of this interface causes this crash, i.e., it cannot execute under the setting of ``Dynamic Shape'', which is one novel feature of MindSpore for dynamically adjusting the shape of layer output. Developers from MindSpore confirmed this bug, and we provided fixed solutions for it, i.e., adopting the specific parameter settings of the interface ``mindspore.ops.interpolate'' to replace it. Developers have accepted this pull request. This bug cannot be detected by existing metamorphic-based methods, as they generate test inputs through parameter and data transformations of single interfaces. They are unable to trigger specifics, such as ``Dynamic Shape'', when optimizing multiple interfaces through their MRs.

\textbf{Accuracy Bugs.}
Among the ten accuracy bugs, two are related to the conversion of different data types (e.g., float64), four are about wrong outputs that do not meet expectations, and the remaining four are related to outliers exposed in the output of specific groups of interfaces.
For example, \tool detects an output error in PyTorch's ``ReLU6'' interface~\cite{accuracybugcase}. While executing the ``ResNet-50'' model for 47 iterations, we observed that its complexity increased excessively. Additionally, the output from the middle layers contains outlier values, such as ``NaN''. Notably, after processing through one ``ReLU6'' interface in the layers of SMR1, the output becomes zero. After reviewing the functionality, we consider an implementation error in PyTorch's ``ReLU6'' when calculating ``NaN'' inputs. After reporting this to the PyTorch community, a developer promptly confirmed our report, screened the bug, and classified it under the ``module: NaNs and Infs'' tag. \myz{The bug-triggering output arises from combinations of inserted structures and SMR1, which existing metamorphic testing-based methods fail to address due to their limited MRs.
}\looseness=-1

Among the 31 detected bugs, 28 of them are detected by our proposed SMRs, and IMRs detect the remaining three. Specifically, all of them are triggered by the IMR, i.e., IMR2, related to the transformation of data type (e.g., the ``Sqrt'' interface of MindSpore does not support int32 data type on CPU). The number of bugs triggered by different SMRs is shown in Table~\ref{tab:mrbugs}. \looseness=-1

\finding{1}{
 \textbf{Answer to RQ2.} Our proposed MRs can trigger: (1) resource and efficiency bugs across multiple interfaces, (2) bugs in loss and gradient calculations, and (3) model structure optimization bugs. Existing methods have detected none of these.
}

\subsection{RQ3: Ablation Study on MRs}
\textbf{Design.} To investigate the effect of our proposed MRs on improving model diversity, we separately apply each SMR to \tool and apply three MRs to \tool while disabling the IMRs for executing 100 rounds. The first kind of variants is denoted as $\tool_{SMR1}$, $\tool_{SMR2}$, $\tool_{SMR3}$, $\tool_{SMR4}$. And the second kind of variants is \myz{denoted} as $\tool_{SMR1,2,3}$, $\tool_{SMR2,3,4}$, $\tool_{SMR1,2,4}$, and $\tool_{SMR1,3,4}$. We compare the diversity of generated models during execution.

\begin{figure}[htbp]
    \centering
    \begin{subfigure}{0.33\textwidth}
        \centering
        \includegraphics[width=\linewidth]{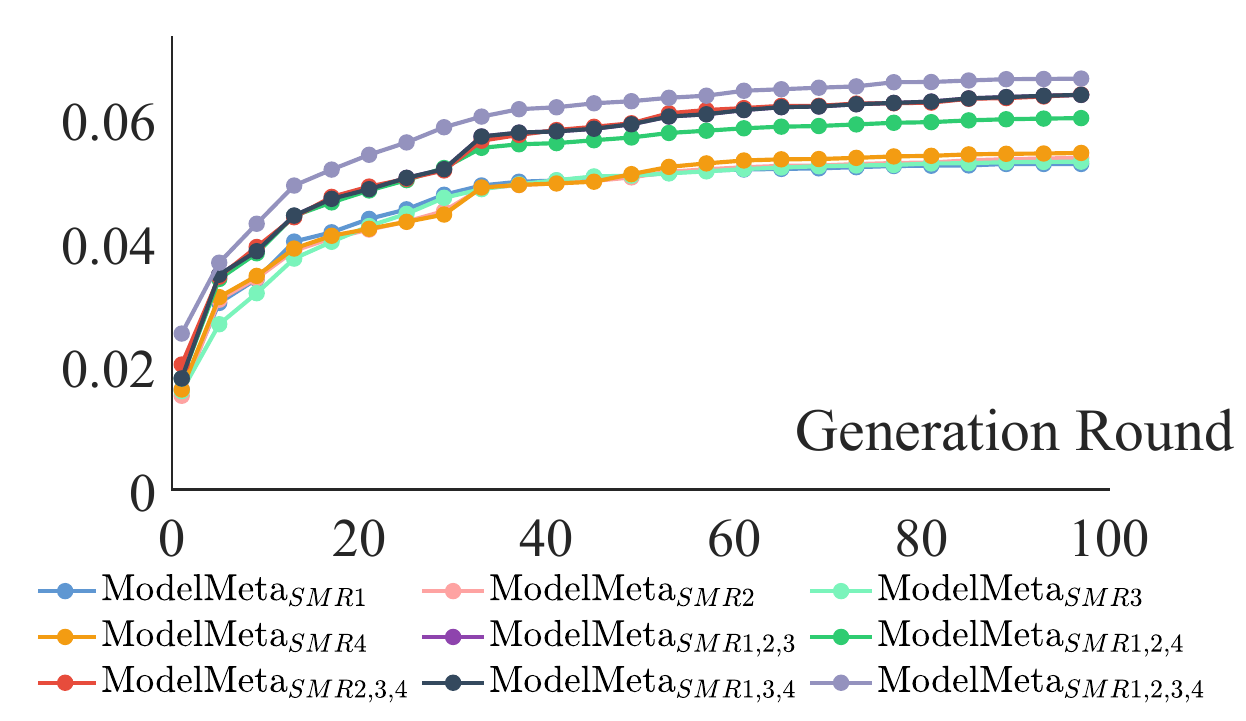}
        
        \caption{LIC}
        \label{fig:rq34mrlic}
    \end{subfigure}
    \begin{subfigure}{0.33\textwidth}
        \centering
        \includegraphics[width=\linewidth]{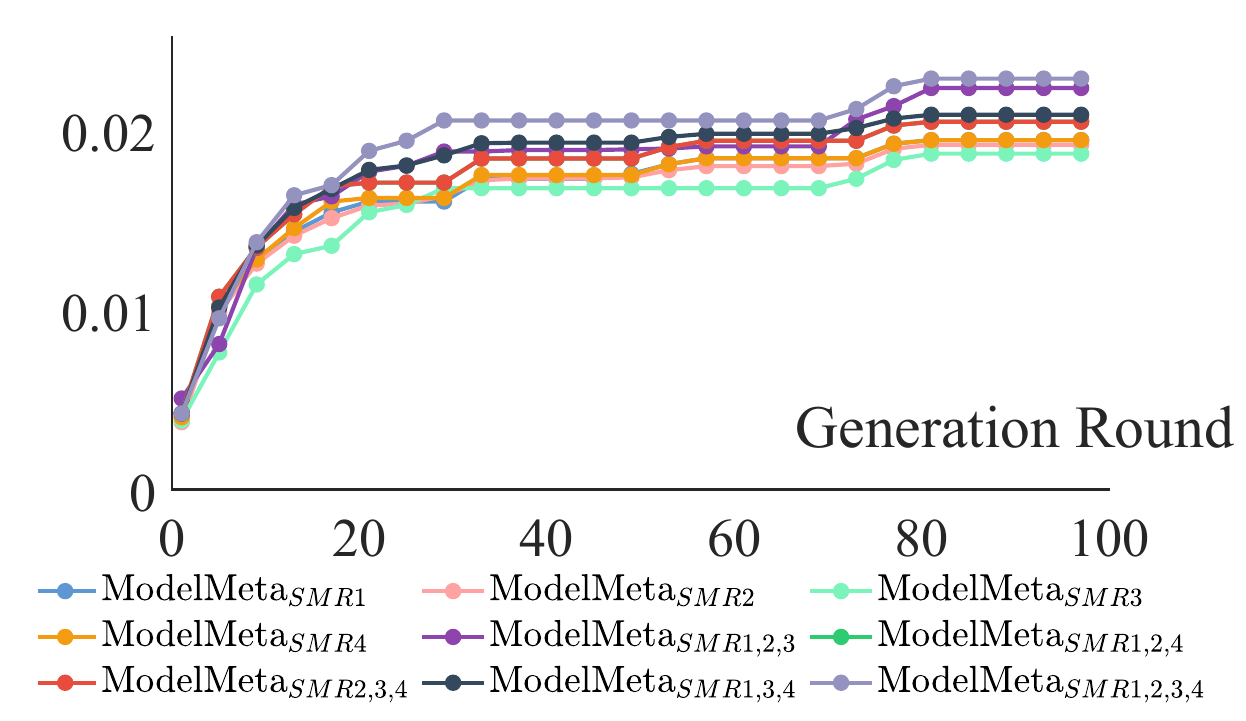}
        
        \caption{LPC}
        \label{fig:rq34mrlpc}
    \end{subfigure}
    \begin{subfigure}{0.32\textwidth}
        \centering
        
        \includegraphics[width=\linewidth]{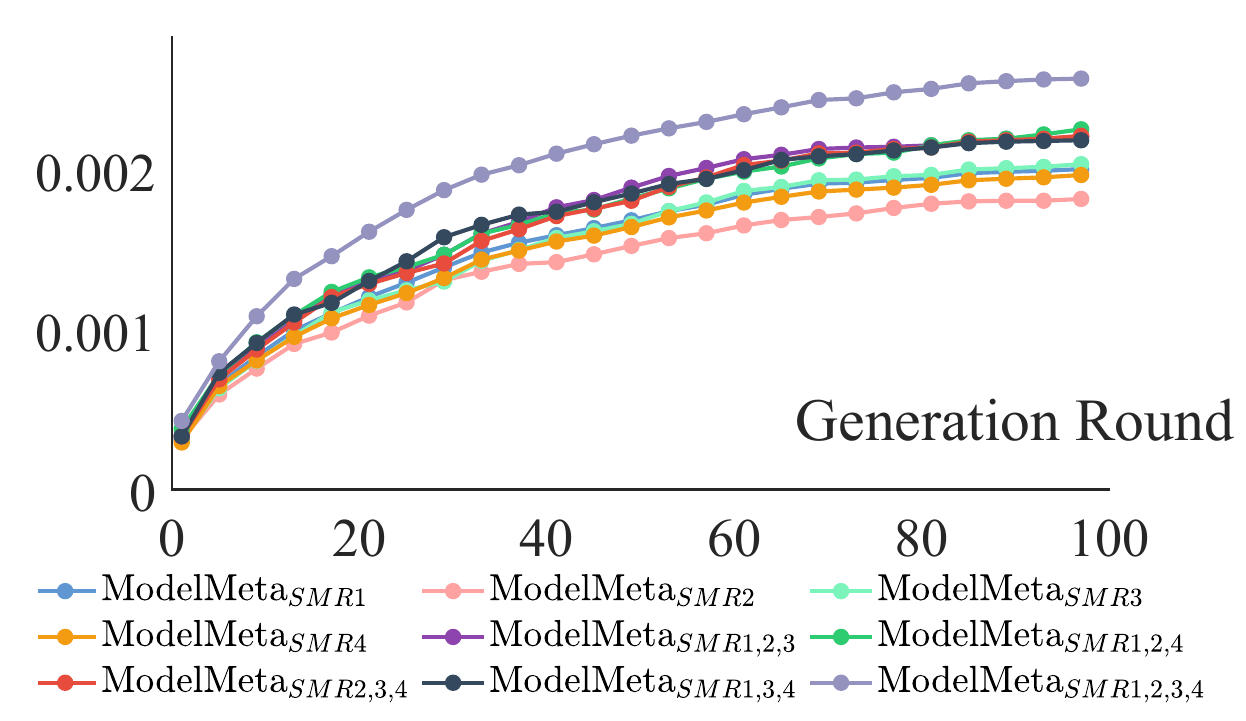}
        \caption{LSC}
        \label{fig:rq34mrlsc}
    \end{subfigure}
    
    \label{fig:rq34}
    \vspace{-4mm}
    \caption{Results about the Ablation Study on Different SMRs}
    \vspace{-4mm}
    
\end{figure}

\textbf{Results about Different SMRs.} Fig.~\ref{fig:rq34mrlic}, ~\ref{fig:rq34mrlpc} and ~\ref{fig:rq34mrlsc} show the results of four SMRs on three diversity metrics, while the $x$ axis represents the execution rounds and the $y$ axis represents the metric values. Experiments show that the original \tool outperforms all variants, achieving the highest values across the three model diversity metrics. The variants $\tool_{SMR1,2,3}$, $\tool_{SMR1,2,4}$, $\tool_{SMR2,3,4}$, and $\tool_{SMR1,3,4}$ perform better than those with single SMR, i.e., $\tool_{SMR1}$, $\tool_{SMR2}$, $\tool_{SMR3}$, and $\tool_{SMR4}$. Specifically, \toolnospace, with the most MRs, achieves higher model diversity earlier. For example, it surpasses other variants around the 18th round on the $LIC$ metric and the 15th round on the $LSC$ metric, ultimately achieving an average improvement of 10.8\% and 9.6\% than other variants, respectively.
Additionally, as shown in Fig. \ref{fig:rq34mrlpc}, variants with more MRs, such as \tool and $\tool_{SMR1,2,3}$, continue generating new diverse models even after other variants with single MRs have converged. From the 74th to the 80th round, \tool and $\tool_{SMR1,2,3}$ achieve further improvements of 11.2\% and 8.2\% on the $LPC$ metric, respectively.
Moreover, variants with the same number of MRs yield similar performance across the metrics, but this does not indicate redundancy. The interface sequences, parameter settings, and input shapes/dimensions differ between variants like $\tool_{SMR1}$, $\tool_{SMR2}$, $\tool_{SMR3}$, and $\tool_{SMR4}$. This highlights how our MRs contribute to increased model diversity by introducing new structures. All our proposed MRs are necessary since they can contribute to different model diversity improvements for \toolnospace. Finally, variants with more MRs cover results from those with fewer MRs, demonstrating the benefits of combining different MRs for greater model structure diversity.

\finding{1}{
 \textbf{Answer to RQ3.} 
    Our proposed MRs can introduce different covered model structures, thereby improving the diversity of models generated by \tool. Moreover, incorporating more MRs into \tool generates even more diverse models.
    }

\subsection{RQ4: Ablation Study on QR-DQN Strategy}
\textbf{Design.} The QR-DQN strategy is designed to guide the selection of the model generation process of \tool. To thoroughly investigate its contribution, we design its three variants with different guideline strategies, i.e., $\tool_{random}$ with random strategy, $\tool_{mcmc}$ with MCMC strategy~\cite{brooks2011handbook} that has been applied in existing methods~\cite{wang2020lemon,li2023comet}, and $\tool_{dqn}$ with DQN strategy~\cite{mnih2015human} that is the general version compared with QR-DQN strategy. We execute \tool and these methods for 100 rounds to compare the diversity of the generated models, execution time, and the count of invalid generated models. \looseness=-1

\begin{figure}[htbp]
    \centering
    \begin{subfigure}{0.33\textwidth}
        \centering
        \includegraphics[width=\linewidth]{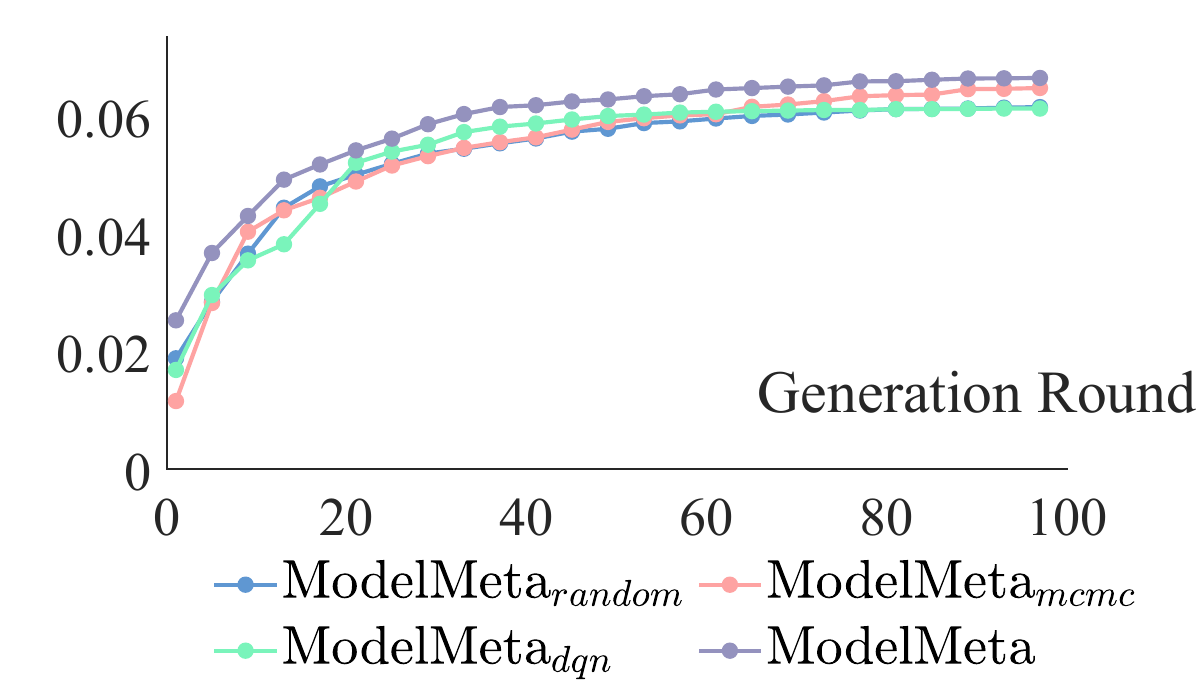}
        
        \caption{LIC}
        \label{fig:rq4lic}
    \end{subfigure}
    \begin{subfigure}{0.33\textwidth}
        \centering
        \includegraphics[width=\linewidth]{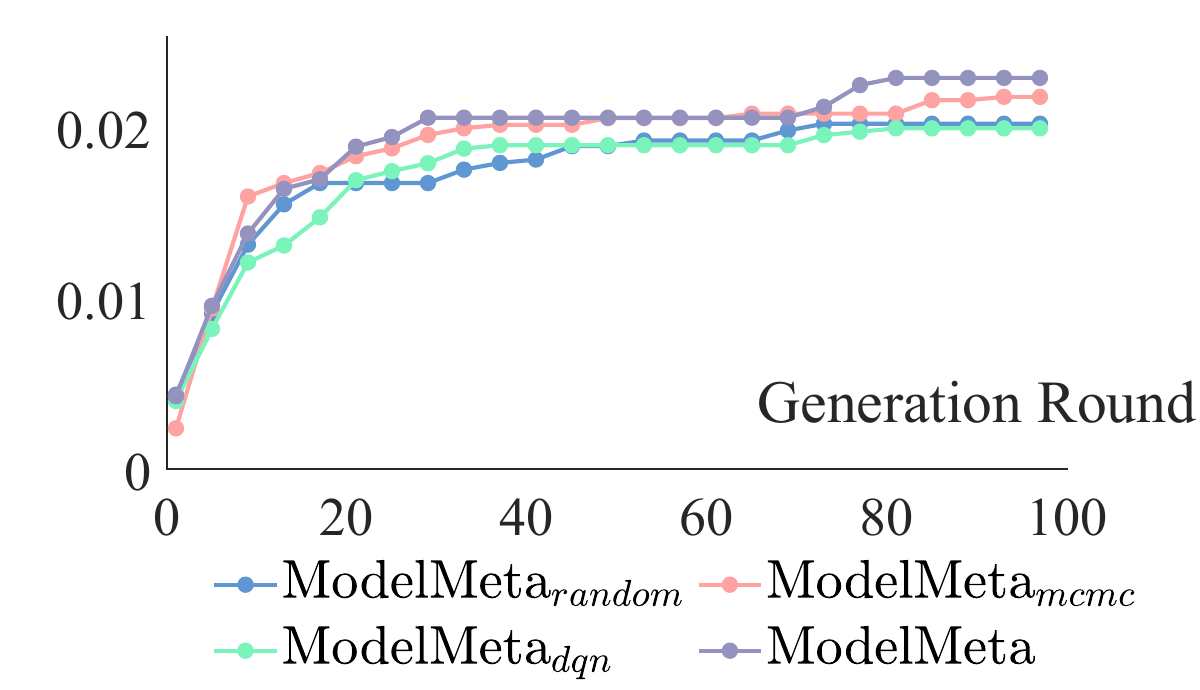}

        \caption{LPC}
        \label{fig:rq4lpc}
    \end{subfigure}
    \begin{subfigure}{0.32\textwidth}
        \centering
        \includegraphics[width=\linewidth]{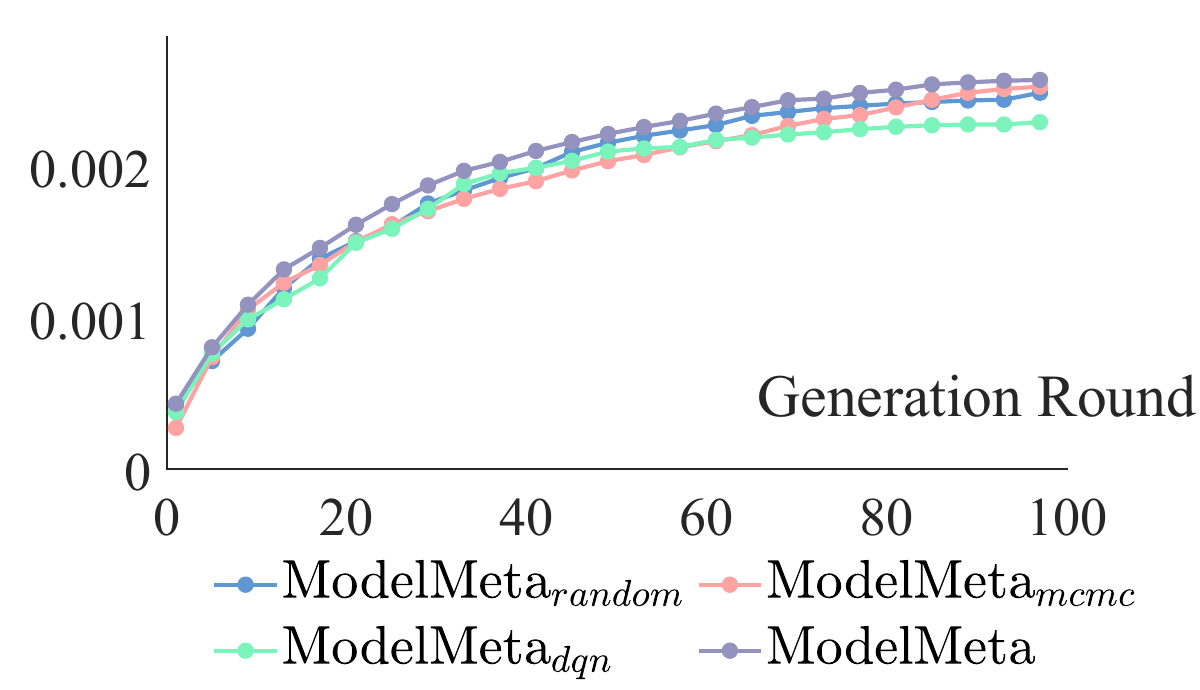}
        
        \caption{LSC}
        \label{fig:rq4lsc}
    \end{subfigure}
    
    \label{fig:rq4}
    \vspace{-4mm}
    \caption{Results about the Ablation Study on Different Guiding Strategies}
    \vspace{-4mm}
\end{figure}

\textbf{Results.} We present the result of three strategies on three diversity metrics in Fig.~\ref{fig:rq4lic},  \ref{fig:rq4lpc} and ~\ref{fig:rq4lsc}. The $x$ axis represents the execution rounds, while the $y$ axis represents the value of different diverse metrics. 
As shown in Fig.~\ref{fig:rq4lic}, \ref{fig:rq4lpc}, and ~\ref{fig:rq4lsc}, \tool achieves higher model diversity more quickly than the other variants, $\tool_{random}$, $\tool_{mcmc}$, and $\tool_{dqn}$ (e.g., \tool surpasses the others starting at the 20th round based on the $LSC$ metric). Additionally, \tool continues generating diverse models even after other variants have converged. For example, \tool improves the $LPC$ metric by 9.3\% from the 70th to 80th round, while the other variants either show continued improvement (e.g., $\tool_{mcmc}$) or only slight improvements (e.g., $\tool_{random}$ and $\tool_{dqn}$ improve by 1.9\% and 3.1\%, respectively).
We find that \tool performs similarly to the other two strategies on simpler models with lower complexity (11 models in our benchmark) based on the three model diversity metrics. Specifically, \tool shows average improvements of 1.4\%, 2.3\%, and 2.8\% over $\tool_{mcmc}$, $\tool_{dqn}$, and $\tool_{random}$, respectively. However, on more complex models (e.g., Deeplabv3, SSD-ResNet50-FPN, Yolov3-DarkNet53, and Yolov4), \tool outperforms both strategies, with average improvements of 16.9\% over $\tool_{mcmc}$, 18.2\% over $\tool_{dqn}$, and 21.2\% over $\tool_{random}$. Additionally, \tool excels on models for new tasks (e.g., GPT-2 and SRGAN), achieving average improvements of 9.4\%, 11.1\%, and 13.8\% over the same variants. This demonstrates \tool's superiority with larger, more complex models and its adaptability to new tasks like image and text generation.\looseness=-1

\myz{Besides, \tool takes the most execution time (17.1 seconds) compared to the other variants: $\tool_{mcmc}$ (15.8 seconds), $\tool_{random}$ (12.6 seconds), and $\tool_{dqn}$ (12.0 seconds). However, the extra time spent is worthwhile, as \tool significantly improves model diversity, as shown by $LIC$, $LPC$, and $LSC$ results. Additionally, \tool generates fewer invalid models (9.4\%) than $\tool_{mcmc}$ (12.6\%), $\tool_{dqn}$ (13.2\%), and $\tool_{random}$ (16.8\%), demonstrating its effectiveness in selecting appropriate MRs for seed models.
}\looseness=-1

\finding{1}{
 \textbf{Answer to RQ4.} 
    \myz{The QR-DQN strategy can generate more valid and diverse models, particularly for complex models, and adapts to new tasks without significantly increasing time.}\looseness=-1
}


\section{Threats to Validity}

\textbf{Internal Threats to Validity.}
It comes from the bias of the \tool implementation, which affects the reliability of our study. To overcome this kind of threat, all authors reviewed the \tool design to ensure theoretical soundness in model generation and bug detection. One author implements it independently, and the other authors perform cross-validation to verify correctness.\looseness=-1

\textbf{External Threats to Validity.} \myz{It arises from the adopted benchmark, which affects the scalability of our study. To address this, we use three popular DL frameworks and 17 models across ten task types from real-world scenarios, covering typical interface combinations and ensuring the generalization ability of \tool. Besides, \tool already achieves high API, code coverage, and test input diversity based on our proposed MRs. Additionally, we provide extension abilities to introduce new diverse structure MRs and diverse groups of interfaces. } \looseness=-1

\textbf{Construct Threats to Validity.}  It mainly comes from the experimental parameter settings and affects the correctness of the experimental result. To address this, we follow recommended defaults from prior studies, including bug detection thresholds and runtime parameters, ensuring accurate reproduction of baseline methods. \toolnospace's execution parameters and detection thresholds are optimized based on large-scale experiments for minimal false positives and high diversity of test inputs.\looseness=-1


\section{Related Work}
\label{sec:relatedwork}

Existing methods adopt the DL models or single interfaces as test inputs and conduct differential testing on their execution results across multiple frameworks to alleviate the challenge of lacking test oracles. However, the performance of such methods is limited by the implementation differences of DL frameworks and version upgrades. Therefore, researchers further apply metamorphic testing, generating new test inputs and detecting bugs based on designed MRs~\cite{wang2022eagle,ding2017validating,chen2024miss}. 
Overall, existing methods can be divided into (1) API-level, (2) model-level, and (3) metamorphic testing-based methods. \looseness=-1

\textbf{API-Level Testing Methods.} 
DL frameworks encapsulate complex algorithms into various interfaces, offering users a common development platform. Therefore, some researchers~\cite{Wei2022FreeLF,zhang2021predoo,zhang2021duo} focus on analyzing these interfaces and examining parameter settings and inputs under constrained conditions to detect bugs. 
Wei et al.~\cite{Wei2022FreeLF} propose Freefuzz, which mines the input of the framework interface from technical communities and wild code fragments. Then, it mutates the collected original inputs and tests the target frameworks by checking the consistency of the execution result across CPU and GPU. 
Zhang et al.~\cite{zhang2021predoo} propose Predoo, a DL operator accuracy testing method that mutates the test input of DL operators to extend the numerical accuracy to maximize the detection of accuracy errors in DL frameworks across CPU and GPU.
Duo~\cite{zhang2021duo} optimizes test input generation and testing by adopting nine mutation operators to evaluate DL framework interfaces on TensorFlow, PyTorch, MNN, and MXNet for detecting bugs.

\textbf{Model-Level Testing Methods.} Since DL models deeply rely on DL frameworks, any framework bugs can adversely affect DL models in their development, execution, and deployment. Therefore, researchers adopt DL models as test inputs for DL frameworks and analyze the model outputs across different frameworks.
CRADLE~\cite{pham2019cradle} is the first method to analyze the output consistency of publicly available models across different Keras~\cite{keras} backends (i.e., TensorFlow, CNTK, Theano). Muffin~\cite{gu2022muffin} designs preset structure templates combined with evolutionary algorithms to generate new models and detect inconsistency bugs during training, leveraging Keras on different backends. LEMON~\cite{wang2020lemon} introduces structure and weight mutations to create diverse models, conducting testing consistent with CRADLE. 
COMET ~\cite{li2023comet} adds two new kinds of mutation operators (parameter and input mutation) and three diversity metrics compared to LEMON, using MCMC~\cite{Andrieu2004AnIT} to guide mutations and performs differential testing between TensorFlow~\cite{tensorflow} and PyTorch~\cite{torch} by third-party tools like TF2ONNX~\cite{TF2ONNX} to convert models. 
Mu et al.~\cite{mu2024devmut} propose DevMuT, which adopts developers' expertise to design mutation operators and constraints for generating new diverse models close to real development. It adopts the Double-Q learning strategy~\cite{hasselt2010double} to guide the mutation and detect bugs among the training and inference process of generated models.
\looseness=-1

\textbf{Metamorphic Testing-based Methods.} 
Wang et al.~\cite{wang2022eagle} propose EAGLE, which generates equivalent test inputs based on 16 MRs for framework interfaces, and then detects bugs by analyzing the consistency of the execution results between equivalent test inputs. 
Chen et al.~\cite{chen2024miss} propose Meta, which can efficiently detect interface implementation errors and accuracy bugs guided with 18 MRs designed for modifying interface parameters and input tensors.
Ding et al.~\cite{ding2017validating} propose 11 MRs related to data transformation from the train, validation, and test set. The MRs require that the classification accuracy of DL models remain consistent after training or inference using the modified data.
\looseness=-1

Compared to existing methods, \tool introduces three key innovations.
(1) Existing MRs focus on input tensor properties (e.g., shape, dimensions) and layer parameters (e.g., value types), targeting accuracy and crash bugs within a single framework interface. In contrast, our MRs leverage structural diversity in test inputs, enabling the detection of bugs arising from combinations of multiple interfaces.
(2) Methods like EAGLE and Meta apply MRs randomly, without utilizing historical execution data, leading to unstable performance and limited test input diversity. In contrast, \tool combines QR-DQN with $\epsilon$-greedy and Thompson Sampling strategies to optimize test input generation, ensuring stable performance by selecting diverse MRs each round and revisiting previous models to explore diverse spaces.
(3) While existing methods focus on single-interface inference results and detect inconsistency and crash bugs, \tool monitors execution time and resource usage to detect abnormal behaviors (e.g., increasing memory usage). This allows it to identify resource bugs (e.g., memory leaks) and efficiency bugs (e.g., excessive execution time due to poor optimization). Additionally, \tool detects crash and accuracy bugs related to incorrect framework optimizations and complex execution stages (e.g., loss calculation and back-propagation during training), highlighting the effectiveness of our MRs in bug detection. \looseness=-1

\section{Conclusion}
\label{sec:conclusion}
Existing DL framework testing methods using metamorphic testing face challenges such as limited MRs, generalization ability, and bug detection types. To address these limitations, we propose \toolnospace, a model-level DL framework metamorphic testing method. \tool incorporates four structure MRs and can introduce diverse interface combinations while integrating interface MRs to enhance test input diversity. It also introduces new test oracles to detect a broader range of bugs, including accuracy, efficiency, crash, and resource bugs. Finally, \tool leverages the QR-DQN strategy combined with the $\epsilon$-greedy and Thompson Sampling strategy to guide test input generation effectively. Our extensive evaluation demonstrates \toolnospace’s promising performance in generating valuable test inputs and detecting new bugs. \looseness=-1

\vspace{-2mm}
\section{Data Availability}
\label{sec:dataavailable}

The executable codes and all the experimental results are available on our website~\cite{sharelink}. We also encourage researchers to contribute new MRs by following our study.

\vspace{-2mm}
\section*{Acknowledgement}
This work is partially supported by the National Natural Science Foundation of China (U24A20337, 62372228), the Shenzhen-Hong Kong-Macau Technology Research Programme (Type C) (Grant No.SGDX20230821091559018), 
the Open Project of State Key
Laboratory for Novel Software Technology at Nanjing University (Grant No. KFKT2024B21) 
and the Fundamental Research Funds for the Central Universities (14380029).

\bibliographystyle{ACM-Reference-Format}
\bibliography{sample-base}

\end{document}